\documentclass[prd,aps,showpacs,nofootinbib,eqsecnum,amsfonts,amsmath,superscriptaddress,
preprintnumbers,a4paper,twocolumn]{revtex4}

\usepackage{epsfig}
\usepackage{graphics}
\usepackage{graphicx}
\usepackage{bm}
\usepackage{color}
\usepackage{amssymb}
\usepackage{psfrag}

\newcommand{\be}{\begin{equation}}
\newcommand{\ee}{\end{equation}}
\newcommand{\ber}{\begin{eqnarray}}
\newcommand{\eer}{\end{eqnarray}}
\newcommand{\bea}{\begin{eqnarray}}
\newcommand{\eea}{\end{eqnarray}}
\newcommand{\ie}{i.e.}
\newcommand{\dt}{{\rm d}t}
\newcommand{\df}{{\rm d}f}
\newcommand{\dtheta}{{\rm d}\theta}
\newcommand{\dphi}{{\rm d}\phi}

\newcommand{\iotahat}{\hat{\iota}}
\newcommand{\phihat}{\hat{\phi}}
\newcommand{\hc}{{\sf h}}

\newcommand{\balpha}{{\bm \alpha}}
\newcommand{\bbeta}{{\bm \psi}}
\newcommand{\fcut}{f_{\rm cut}}
\newcommand{\fmerg}{f_{\rm merg}}
\newcommand{\fring}{f_{\rm ring}}
\newcommand{\rmi}{{\rm i}}
\newcommand{\matA}{{\sf A}}
\newcommand{\vecB}{{\bf B}}
\newcommand{\Mchirp}{M_{\rm c}}
\newcommand{\veclambda}{{\bm \lambda}}
\newcommand{\vectheta}{{\bm \theta}}

\def\qnr{{\textrm{\mbox{\tiny{QNR}}}}}

\begin{document}

\title{A template bank for gravitational waveforms from coalescing binary black holes: \\ non-spinning binaries}

\author{P.~Ajith}
\affiliation{Max-Planck-Institut f\"ur Gravitationsphysik 
(Albert-Einstein-Institut) and Leibniz Universit\"at Hannover, 
Callinstr.~38, 30167~Hannover, Germany}

\author{S.~Babak}
\affiliation{Max-Planck-Institut f\"ur Gravitationsphysik (Albert-Einstein-Institut), 
Am~M\"uhlenberg 1, 14476~Golm, Germany}

\author{Y.~Chen}
\affiliation{Max-Planck-Institut f\"ur Gravitationsphysik (Albert-Einstein-Institut), 
Am~M\"uhlenberg 1, 14476~Golm, Germany}

\author{M.~Hewitson}
\affiliation{Max-Planck-Institut f\"ur Gravitationsphysik 
(Albert-Einstein-Institut) and Leibniz Universit\"at Hannover, 
Callinstr.~38, 30167~Hannover, Germany}

\author{B.~Krishnan}
\affiliation{Max-Planck-Institut f\"ur Gravitationsphysik (Albert-Einstein-Institut), 
Am~M\"uhlenberg 1, 14476~Golm, Germany}

\author{A.~M.~Sintes}
\affiliation{Max-Planck-Institut f\"ur Gravitationsphysik (Albert-Einstein-Institut), 
Am~M\"uhlenberg 1, 14476~Golm, Germany}
\affiliation{Departament de F\'{\i}sica, Universitat de les Illes
Balears, Cra. Valldemossa Km. 7.5, E-07122 Palma de Mallorca, Spain}

\author{J.~T.~Whelan}
\affiliation{Max-Planck-Institut f\"ur Gravitationsphysik (Albert-Einstein-Institut), 
Am~M\"uhlenberg 1, 14476~Golm, Germany}

\author{B.~Br\"ugmann}
\affiliation{Theoretisch-Physikalisches Institut, Friedrich Schiller Universit\"at Jena,
Max-Wien-Platz 1, 07743~Jena, Germany}

\author{P.~Diener}
\affiliation{Center for Computation \& Technology, Louisiana State University, Baton Rouge, LA, USA}
\affiliation{Department of Physics and Astronomy, Louisiana State University, Baton Rouge, LA, USA}

\author{N.~Dorband}
\affiliation{Max-Planck-Institut f\"ur Gravitationsphysik (Albert-Einstein-Institut), 
Am~M\"uhlenberg 1, 14476~Golm, Germany}

\author{J.~Gonzalez}
\affiliation{Theoretisch-Physikalisches Institut, Friedrich Schiller Universit\"at Jena,
Max-Wien-Platz 1, 07743~Jena, Germany}
\affiliation{Instituto de F\a'{\i}sica y Matem\a'aticas,
Universidad Michoacana de San Nicol\a'as de Hidalgo, Edificio C-3,
Cd. Universitaria. C. P. 58040 Morelia, Michoac\a'an, M\a'exico}

\author{M.~Hannam}
\affiliation{Theoretisch-Physikalisches Institut, Friedrich Schiller Universit\"at Jena,
Max-Wien-Platz 1, 07743~Jena, Germany}

\author{S.~Husa}
\affiliation{Theoretisch-Physikalisches Institut, Friedrich Schiller Universit\"at Jena,
Max-Wien-Platz 1, 07743~Jena, Germany}

\author{D.~Pollney}
\affiliation{Max-Planck-Institut f\"ur Gravitationsphysik (Albert-Einstein-Institut), 
Am~M\"uhlenberg 1, 14476~Golm, Germany}

\author{L.~Rezzolla}
\affiliation{Max-Planck-Institut f\"ur Gravitationsphysik (Albert-Einstein-Institut), 
Am~M\"uhlenberg 1, 14476~Golm, Germany}

\author{L.~Santamar\'ia}
\affiliation{Theoretisch-Physikalisches Institut, Friedrich Schiller Universit\"at Jena,
Max-Wien-Platz 1, 07743~Jena, Germany}
\affiliation{Max-Planck-Institut f\"ur Gravitationsphysik (Albert-Einstein-Institut), 
Am~M\"uhlenberg 1, 14476~Golm, Germany}

\author{U. Sperhake}
\affiliation{Theoretisch-Physikalisches Institut, Friedrich Schiller Universit\"at Jena,
Max-Wien-Platz 1, 07743~Jena, Germany}

\author{J.~Thornburg}
\affiliation{Max-Planck-Institut f\"ur Gravitationsphysik (Albert-Einstein-Institut), 
Am~M\"uhlenberg 1, 14476~Golm, Germany}
\affiliation{School of Mathematics, University of Southampton, Southampton SO17~1BJ, England}

\bigskip

\date{\today}

\begin{abstract}
Gravitational waveforms from the inspiral and ring-down stages of the
binary black hole coalescences can be modelled accurately by
approximation/perturbation techniques in general relativity. Recent
progress in numerical relativity has enabled us to model also the
non-perturbative merger phase of the binary black-hole coalescence
problem. This enables us to \emph{coherently} search for all three
stages of the coalescence of non-spinning binary black holes using a
single template bank. Taking our motivation from these results, we
propose a family of template waveforms which can model the inspiral,
merger, and ring-down stages of the coalescence of non-spinning binary
black holes that follow quasi-circular inspiral. This two-dimensional
template family is explicitly parametrized by the physical parameters
of the binary. We show that the template family is not only
\emph{effectual} in detecting the signals from black hole
coalescences, but also \emph{faithful} in estimating the parameters of
the binary. We compare the sensitivity of a search (in the context of
different ground-based interferometers) using all three stages of the
black hole coalescence with other template-based searches which look
for individual stages separately. We find that the proposed search is
significantly more sensitive than other template-based searches for a
substantial mass-range, potentially bringing about remarkable
improvement in the event-rate of ground-based interferometers. As part
of this work, we also prescribe a general procedure to construct
interpolated template banks using non-spinning black hole waveforms
produced by numerical relativity.
\end{abstract}

\preprint{LIGO-P070111-00-Z}
\preprint{AEI-2007-143}

\maketitle

\section{Introduction}
\label{sec:intro}

A network of ground based gravitational-wave (GW) detectors (LIGO,
Virgo, GEO~600, TAMA) is currently collecting data, which a world-wide
scientific collaboration is involved in analyzing.  Among the most
promising sources detectable by these observatories are coalescing
compact binaries consisting of black holes (BHs) and/or neutron stars
spiraling toward each other as they lose orbital energy and angular
momentum through gravitational-wave emission.  The gravitational-wave
signal from coalescing binaries is conventionally split into three
parts: inspiral, merger and ring down.  In the first stage, the two
compact objects, usually treated as point masses, move in
quasi-circular orbits (eccentricity, if present initially, is quickly
radiated away). This part of the waveform is described very well by
the post-Newtonian (PN) approximation of general relativity. In this
approximation the Einstein equations are solved in the near zone
(which contains the source) using an expansion in terms of the (small)
velocity of the point masses. In the far zone, the vacuum equations
are solved assuming weak gravitational fields, and these two solutions
are matched in the intermediate
region~\cite{fock,Blanchet:1985sp,Wagoner:1976am}.

The PN approximation breaks down as the two compact objects approach
the ultra-relativistic regime and eventually merge with each other.
Although various resummation methods, such as Pad\'e~\cite{DIS98} and
effective-one-body (EOB) approaches~\cite{Buonanno:1998gg}, have been
developed to extend the validity of the PN approximation, unambiguous
waveforms in the merger stage must be calculated numerically in full
general relativity. Recent breakthroughs in numerical relativity 
\cite{Pretorius:2005gq,Campanelli:2005dd,Baker05a} have allowed many groups
\cite{Pretorius:2005gq,Campanelli:2005dd,Baker05a,Herrmann2006,
Sperhake2006,Bruegmann:2006at,Thornburg-etal-2007a,Etienne:2007hr}
to evolve BH binaries fully numerically for the last several orbits
through the plunge to single BH formation. The field is now rapidly
developing the capability to routinely evolve generic black-hole
binary configurations in the comparable-mass regime, and to accurately
extract the gravitational-wave signal. Important milestones include
simulations of unequal-mass binaries and calculations of the gravitational 
recoil effect and the evolution of black-hole binaries with spin 
\cite{Baker2006b,Gonzalez06tr,Campanelli2006b,Campanelli2006c,Herrmann:2007ac,
Koppitz-etal-2007aa,Gonzalez:2007hi,Campanelli2007,Campanelli:2006fy,
Pollney:2007ss,Rezzolla-etal-2007}.

Comparisons with post-Newtonian results are essential for data analysis
efforts, and several groups have published results showing good agreement of
various aspects of non-spinning simulations with post-Newtonian predictions
(see e.g.~\cite{Baker:2006yw,Baker:2006ha,Buonanno:2006ui,Berti:2007fi,
Pan:2007nw,Ajith:2007qp,Hannam:2007ik,Boyle:2007ft}),
and first results for certain configurations with spin
have also become available \cite{Brugmann:2007zj,Lousto:2007db}. 
In order to overcome phase inaccuracies
in long evolutions, significant progress has been made by the Caltech-Cornell
group using spectral codes \cite{Scheel-etal-2006:dual-frame,Pfeiffer:2007},
and by the Jena group using higher (sixth) order finite differencing
\cite{Husa2007a}. Methods to reduce the eccentricity to around $10^{-3}$ (so far
only for equal-mass binaries) have been presented by the Caltech-Cornell group
\cite{Pfeiffer:2007}, and the Jena group \cite{Husa:2007ec} (using
initial parameters from PN solutions that take into account radiation
reaction). Current numerical waveforms can be generated for the last ($\lesssim 10$)
orbits, and these waveforms can be joined continuously with analytic PN 
inspiral waveforms to obtain one full signal. This was done in
\cite{Buonanno:2006ui,Pan:2007nw,Ajith:2007qp,Buonanno:2007pf}.
Indeed, there are no fundamental obstructions to generating the whole
waveform, including long inspiral over hundreds of orbits, by solving
the full Einstein equations numerically. But, not only would this be
computationally prohibitive with current methods, it is also
unnecessary: the PN formalism is known to work very well in the
weak-field regime (when the BHs are well-separated), and is a low-cost
and perfectly adequate substitute to fully general relativistic
solutions in that regime.

The numerically generated part of the gravitational-wave signal from
coalescing binaries also includes the final stage of the coalescence,
when a single perturbed black hole is formed and it rapidly loses its
deviations from a Kerr black hole via gravitational waves. This part
of the signal can be decomposed as a superposition of exponentially
damped modes, and is called quasi-normal mode `ring down', by analogy
with the vibrations of a bell. The detectable part of the ring down is
rather short and only a few modes (if not only the dominant one) are
expected to be important/detectable by initial ground-based
observatories. This will not be true, however, for the advanced
detectors \cite{Berti:2007zu} and certainly it is not the case for
LISA, the planned space-borne gravitational-wave observatory. Indeed,
the majority of the signal-to-noise ratio (SNR) comes from the
quasi-normal mode ringing of binary systems with a total mass above a
few $10^6\, M_{\odot}$ \cite{Berti:2005ys}.  For LISA, and also
perhaps for the next generation of ground based detectors, it will be
possible to detect several quasi-normal modes and test the `no hair'
theorem, according to which all modes are functions of a BH's mass and
spin~\cite{Dreyer:2003bv,Berti:2005qd, Berti:2005ys}.

Joining analytically modeled inspiral with numerically generated
merger and ring down allows us to produce the complete
gravitational-wave signal from coalescing binaries, and to use it in
the analysis of detector data. There are several benefits to using the
whole signal in searches. The most obvious one is the increase in SNR
in a fully coherent matched filtering search~\cite{FlanHugh98,BuonD00,
DIS01,Ajith:2007qp}.
Increase in SNR implies increase in the event rate and improvement in
the parameter estimation. Including the inspiral, merger and ring down
parts in a template waveform also means that the waveform has a more
complex structure. This extra complexity will also bring about some
improvement in the parameter estimation~\cite{ParamEstim} and possibly
also a reduction in the false alarm rate in analysis of the data from
the ground-based network of detectors.  This is because it is in
general harder for the noise to mimic a complex signal\footnote{At
least we expect this to happen for those binaries for which both the
inspiral and the merger contribute significantly to SNR.}.  For LISA,
the detection of inspiralling super-massive black holes is not a
problem; the SNR is expected to be so large that we expect some
signals to be visible by eye in LISA data.  However, using the full
signal for LISA data analysis is equally important because the full
signal is essential in estimating parameters of the binary with the
required accuracy. This is important not only from the astrophysical
point of view, but also because we need to subtract loud signals from
the data in order to detect/analyze other signals.  Imperfect signal
removal due to errors in the parameter estimation will result in large
residuals and will adversely affect subsequent analyses. Improved 
parameter estimation will also enable GW observations (in conjunction
with electromagnetic observations) to constrain important
cosmological parameters, most importantly the equation of state of dark 
energy~\cite{Schutz86,Markovic:1993cr,ChernoffFinn:1993,HolzHugh05,
Arun:2007hu,ParamEstim}.

The numerical waveforms described above are still computationally
expensive and cannot be used directly to densely cover the parameter
space of the binary BHs that will be searched over by matched
filtering techniques. A promising alternative is to use the
post-Newtonian and numerical-relativity waveforms to construct an
analytic model that sufficiently accurately mimics a true signal
\cite{Ajith:2007qp, Buonanno:2007pf}. In~\cite{Ajith:2007qp} we have
suggested a phenomenological family of waveforms which can match
physical signals from non-spinning binaries in quasi-circular orbits
with fitting factors above 99\%.  In this paper we extend this
formulation to propose a two-parameter family of template waveforms
which are explicitly parametrized by the physical parameters of the
binary. We show that this two-dimensional template family is not only
\textit{`effectual'} in detecting the signals from binary BH
coalescences, but also \textit{`faithful'} in estimating the
parameters of the binary. This family of template waveforms can be
used to densely cover the parameter space of the binary, thus avoiding
the computational burden of generating numerical waveforms in each
grid point in the parameter space. We compute the effectualness and
faithfulness (see Section~\ref{sec:DA} for definitions) of the
template family in the context of three different ground-based
detectors: namely, Initial LIGO, Virgo and Advanced LIGO. We also
compare the sensitivity of a search which coherently includes all
three (inspiral, merger and ring down) stages of the BH coalescence
with other template-based searches which look for each stage
separately.

Our `target signals' are constructed by matching the
numerical-relativity waveforms to a particular family (\emph{TaylorT1}
approximant~\cite{DIS01}) of post-Newtonian waveforms, but this choice
is by no means necessary. Indeed, we expect that more robust ways of
constructing post-Newtonian approximants, such as the effective
one-body approach~\cite{Buonanno:1998gg} or Pad\'e resummation
approach~\cite{DIS98}, will give better agreement with numerical-relativity (NR) waveforms.
But the purpose of the current paper is to explicitly prescribe a
general procedure to produce hybrid and phenomenological waveforms,
and to construct interpolated template banks using parametrized
waveforms.  We show that, given the number of numerical wave cycles we
employ, even a simple PN choice like TaylorT1 leads to very faithful
and effectual templates, and significantly increases the possible
range of gravitational-wave searches. The use of improved PN
approximants will require a smaller number of NR cycles, thereby
further reducing computational cost for template construction.  There
are also other approaches for comparing analytic and numerical
waveforms and for constructing hybrid waveforms (see, for example \cite{Pan:2007nw}); it
would be interesting to compare the results presented in this work
with other approaches presented in the literature.

The paper is structured as follows. In Section~\ref{sec:numrelintro}
we summarize the methods of current numerical-relativity simulations,
including a setup of the initial data that allows an unambiguous
comparison with post-Newtonian results, and the wave extraction
techniques. In Section~\ref{sec:DA} we briefly outline the waveform
generation using the \emph{restricted} post-Newtonian
approximation. There we briefly introduce the main data-analysis
techniques and define notations that are used in the subsequent
sections.  In Section~\ref{sec:phenomtemplates} we construct a
phenomenological template family parametrized only by the masses of
the two individual black holes. First we combine restricted 3.5PN
waveforms \cite{BDEI04} with results from NR simulations to construct
`hybrid' waveforms for the quasi-circular inspiral of non-spinning
binaries with possibly unequal masses. Then, we introduce a
phenomenological family of templates constructed in the frequency
domain. Initially the template family is parametrized by 10
phenomenological parameters. We then find a unique mapping of these 10
parameters to the two physical parameters: namely, the total mass $M$
and the symmetric mass ratio $\eta \equiv M_1 M_2 / M^2$, so that the
template family is just two-dimensional. The resulting templates have
remarkably high fitting factors with target waveforms. Here we also
compute the faithfulness of the templates and the bias in the
estimation of the parameter of the binary. A comparison of the
sensitivity of the search using the proposed template family with
other existing template-based searches is also presented. Finally, we
summarize our main results in Section~\ref{sec:summary}.  Some details
of the calculations involved are described in
Appendices~\ref{app:fitfactor} and~\ref{app:horDist}. We adopt
geometrical units throughout this paper: $G=c=1$.

\section{Numerical simulations and wave extraction}
\label{sec:numrelintro}

Numerical simulations were performed with the BAM
\cite{Bruegmann:2006at} and CCATIE \cite{Pollney:2007ss} codes. Both
codes evolve black-hole binaries using the `moving-puncture'
approach \cite{Campanelli:2005dd,Baker05a}.  The method involves
setting up initial data containing two black holes via a
Brill-Linquist-like wormhole construction \cite{Brill:1963yv}, where
the additional asymptotically flat end of each wormhole is
compactified to a point, or `puncture'.  A coordinate singularity
exists at the puncture, but can be stably evolved using standard
finite-difference techniques, and is protected by causality from
adversely affecting the physically relevant external spacetime. This
prescription allows black holes to be constructed on a 3D Cartesian
numerical grid without recourse to excision techniques, and also
provides a simple way to generate any number of moving, spinning black
holes \cite{Bowen80,Brandt97b}. Given an initial configuration of two
black holes, the data are evolved using a conformal and traceless
`3+1' decomposition of Einstein's equations
\cite{Nakamura87,Shibata95,Baumgarte99}. In addition the gauge is
evolved using the `1+log'~\cite{Bona94b,Alcubierre02a} and
`$\Gamma$-driver' equations \cite{Alcubierre00a,Alcubierre02a} and
the coordinate singularity in the conformal factor is dealt with by
evolving either the regular variable $\chi = \psi^{-4}$
\cite{Campanelli:2005dd} (in BAM) or $\phi = \ln \psi$ (in CCATIE),
which diverges `slowly' enough so as not lead to numerical
instabilities.  The standard moving puncture approach consists of all
these techniques, and causes the `punctures' to quickly assume a
cylindrical asymptotics \cite{Hannam:2006vv}, and allows them to move
across the numerical grid. This method has been found to allow
accurate, stable simulations of black holes over many ($>10$) orbits
through merger and ring down.

In the initial data construction we must specify the masses, locations
and momenta of the two black holes (we do not consider spinning black
holes in this work).  The mass of each black hole, $M_i$, is specified
in terms of the Arnowitt-Deser-Misner (ADM) mass at each puncture.
This corresponds to the mass at the other asymptotically flat end
which is, to a very good approximation, equal to irreducible mass of
the apparent-horizon mass
\cite{Schnetter:2006yt,Dennison:2006nq,Tichy:2003qi}
\begin{equation} M_i = \sqrt{
    \frac{A_i}{16 \pi} }\,.
\end{equation}
where $A_i$ is the area of the apparent horizon.  We assume that this
mass is the same as the mass used in post-Newtonian formulas.  This
assumption is really expected to be true only in the limit where the
black holes are infinitely far apart and stationary. As such we
consider any error in this assumption as part of the error due to
starting the simulation at a finite separation. The important point is
that a binary with horizon masses $M_1$ and $M_2$ should be compared
with a post-Newtonian system with the same mass parameters.  This
allows us to provide the same overall scale $M = M_1 + M_2$ for both
numerical and post-Newtonian waveforms, and is crucial for comparison
and matching.

The initial momenta of the black holes are chosen to correspond approximately to 
quasi-circular (low eccentricity) inspiral. For equal-mass evolutions performed
with the CCATIE code, parameters for quasi-circular orbit were determined
by minimizing an effective potential for the binary~\cite{Cook94,Baumgarte00a,Pollney:2007ss}. 
For the unequal-mass simulations 
 performed with the BAM code \cite{Gonzalez06tr}, initial momenta were specified 
 by the 3PN-accurate quasi-circular formula given in Section~VII of \cite{Bruegmann:2006at}. 
 For the longer unequal-mass simulations performed with higher-order spatial 
 finite-difference methods \cite{Husa2007a} and used for verification, the initial 
 momenta were taken from a PN prescription that takes radiation reaction into account 
 to reduce the initial eccentricity to below $e \approx 10^{-3}$ \cite{Husa:2007ec}. 

The Einstein equations are solved numerically with standard
finite-difference techniques.  Spatial derivatives are calculated at
fourth- or sixth-order accuracy, and the time evolution is performed
with a fourth-order Runge-Kutta integration. Mesh refinement is used to
achieve high resolution around the punctures and low resolutions far
from the black holes, allowing the outer boundary to be placed very
far (at least $>300M$) from the sources. Full details of the numerical
methods used in the two codes are given in \cite{Bruegmann:2006at} for
BAM and \cite{Pollney:2007ss} for CCATIE.

In the wave-zone, sufficiently far away from the source, the spacetime
metric can be accurately described as a perturbation of a flat
background metric. Let $h_{ab}$ denote the metric perturbation where
$a,b$ denote spacetime indices, and $t$ be the time coordinate used in
the numerical simulation to foliate the spacetime by spatial slices.
Working in the transverse-traceless (TT) gauge, all the information
about the radiative degrees of freedom is contained in the spatial
part $h_{ij}$ of $h_{ab}$, where $i,j$ denote spatial indices.  Let us
use a coordinate system $(x,y,z)$ on a spatial slice so that the
$z$-axis is parallel to the total angular momentum of the binary
system at the starting time.  Let $\iota$ be the inclination angle
from the $z$-axis, and let $\phi$ be the phase angle and $r$ the
radial distance coordinates so that $(r,\iota,\phi)$ are standard
spherical coordinates in the wave-zone.

The radiative degrees of freedom in $h_{ab}$ can be written in terms of two
polarizations $h_+$ and $ h_\times$:
\begin{equation}
  \label{eq:2}
  h_{ij}  = h_+ (\mathbf{e}_{+})_{ij} +
  h_\times(\mathbf{e}_\times)_{ij},
\end{equation}
where $\mathbf{e}_{+,\times}$ are the basis tensors for
transverse-traceless tensors in the wave frame
\begin{equation}
  \label{eq:9}
  (\mathbf{e}_+)_{ij} = \iotahat_i \iotahat_j -
  \phihat_i \phihat_j\,, \quad \textrm{and} \quad 
  (\mathbf{e}_\times)_{ij} = \iotahat_{i}\phihat_{j} + \iotahat_{j}\phihat_{i}\,.
\end{equation}
Here $\iotahat$ and $\phihat$ are the unit vectors in the $\iota$ and
$\phi$ directions, respectively, and the wave propagates in the radial
direction.

In our numerical simulations, the gravitational waves are extracted by
two distinct methods. The first one uses the Newman-Penrose Weyl
tensor component $\Psi_4$~\cite{Newman62a,Stewart:1990uf} which is a
measure of the outgoing transverse gravitational radiation in an
asymptotically flat spacetime.  In the wave-zone it can be written in
terms of the complex strain $\hc = h_+ - \mathrm{i} h_\times$
as~\cite{Teukolsky73},
\begin{equation}
  \hc = \lim_{r\rightarrow\infty} \int^t_0 \dt^\prime
  \int^{t^\prime}_0 \dt^{\prime\prime} \Psi_4.  \label{eq:psi4}
\end{equation}
An alternative method for wave extraction determines the waveform 
via gauge-invariant
perturbations of a background Schwarzschild spacetime, via the
Zerilli-Moncrief formalism (see~\cite{Nagar:2005ea} for a review). In
terms of the even ($Q^{+}_{\ell m}$) and odd ($Q^{\times}_{\ell
  m}$) parity master functions, the gravitational wave strain
amplitude is then given by
\begin{equation}
 \hc = \frac{1}{\sqrt{2}r}
  \sum_{\ell,m}\left(Q^+_{\ell m} - \mathrm{i}
    \int^t_{-\infty}Q^\times_{\ell m}(t^\prime)\dt^\prime\right)
    Y^{-2}_{\ell m} + \mathcal{O}\left(\frac{1}{r^2}\right).
  \label{eq:zerilli}
\end{equation}

Results from the BAM code have used the Weyl tensor component $\Psi_4$
and Eq. (\ref{eq:psi4}), with the implementation described
in~\cite{Bruegmann:2006at}. While the CCATIE code computes waveforms
with both methods, the AEI-CCT waveforms used here were computed
using the perturbative extraction and Eq. (\ref{eq:zerilli}).  Beyond
an appropriate extraction radius (that is, in the wave-zone), the two
methods for determining $\hc$ are found to agree very well for
moving-puncture black-hole evolutions of the type considered
here~\cite{Koppitz-etal-2007aa}.

It is useful to discuss gravitational radiation fields in terms of
spin-weighted $s=-2$ spherical harmonics $Y^{s}_{\ell m}$, which represent symmetric
tracefree 2-tensors on a sphere, and in this
paper we will only consider the dominant $\ell=2,\ m=\pm 2$ modes (see
\cite{Berti:2007fi} for the higher $\ell$ contribution in the unequal-mass
 case), with basis functions
\begin{eqnarray}
  Y^{-2}_{2-2} & \equiv & \sqrt{\frac{5}{64\pi}} \left(1 -\cos \iota \right)^2
       e^{-2 \mathrm{i}\phi}, \nonumber \\
  Y^{-2}_{22} & \equiv & \sqrt{\frac{5}{64\pi}} \left( 1 +\cos \iota \right)^2
       e^{2 \mathrm{i}\phi}.
\end{eqnarray}
Our `input' numerical relativity waveforms thus correspond to the projections
\begin{equation}
\hc_{\ell m} \equiv \langle Y^{-2}_{\ell m}, \hc \rangle = 
	\int_0^{2\pi} \dphi \int_0^{\pi}
      \hc \, \overline{Y^{-2}_{\ell m}}\, \sin \theta\,\dtheta\,
      \label{eq: scalar_product},
\end{equation}
of the complex strain $\hc$, where the bar denotes complex
conjugation.  In the cases considered here, we have equatorial
symmetry so that $\hc_{22} = \overline{ \hc_{2-2} }$, and
\begin{equation}
  \hc(t) = \sqrt{\frac{5}{64\pi}} e^{2 \mathrm{i}\phi} \left(
    \left(1 + \cos \iota \right)^2       \hc_{22}(t)
    + \left(1 - \cos \iota \right)^2  \bar \hc_{22}(t) \right). 
\label{eq:NRh}
\end{equation}
In this paper, we assume that the binary is optimally-oriented, so that $\iota=0$. Thus
\begin{equation}
\hc(t) = 4 \, \sqrt{\frac{5}{64\pi}} \, \hc_{22}(t) \approx 0.6308\, \hc_{22}(t).
\label{eq:NRhOptOriented}
\end{equation}

\section{Post-Newtonian waveforms and introduction to data-analysis concepts}
\label{sec:DA}

In this Section we will introduce notation that will be used later in
the paper and describe briefly the main data-analysis techniques
currently used in gravitational wave astronomy.

\subsection{Restricted post-Newtonian waveforms}

We use the restricted  PN waveform at mass-quadrupole
order, which has a phase equal to twice the orbital phase up to
highest available order in the adiabatic approximation, and amplitude
accurate up to leading order.  The corresponding $\hc$ is given
by 
\begin{equation}
\hc=\frac{\eta M}{r}v^2(t) e^{2\rmi\phi}\left[(1+\cos\iota)^2 e^{-\rmi \varphi(t)} +
(1-\cos\iota)^2 e^{\rmi \varphi(t)}\right]
\label{eq:PNhOptOrient}
\end{equation}
where $M \equiv M_1+ M_2$ is the total mass, $\eta \equiv M_1 M_2/M^2$ is 
the symmetric mass ratio, $r$ is the observation radius, $\iota$ is the 
inclination angle;  the quantity $v(t)$ 
is an expansion parameter, defined by $v=(M \dot{\varphi}/2)^{1/3}$ 
with $\varphi(t)$ equal to twice the adiabatic orbital phase.  The {\it waveform} 
seen by the detector is given  by  
\be
s(t) = 4\,\eta\, \frac{M}{r} A \, v^2(t) \cos[\varphi(t)+\varphi_0],
\ee
where, for short-lived signals (i.e., with duration much shorter than the earth 
rotation time, as well as de-phasing time scale due to Doppler shifts induced by 
earth motion and rotation),  $A$ and $\varphi_0$ are numerical constants
depending on the relative position and orientation of the source relative to
the detector, as well as the antenna pattern functions of the detector.  In  
PN theory, the adiabatic phase $\varphi(t)$ is determined by the following ordinary 
differential equations (also called the \emph{phasing formula}):
\begin{equation}
\frac{{\rm d}\varphi}{\dt} = \frac{2v^3}{M},\ \ \ \ 
\frac{{\rm d}v}{\dt} = -\frac{{\cal F}(v)}{ME'(v)}.
\label{eq:phasing1}
\end{equation}
In these expressions, $E'(v)={\rm d}E(v)/{\rm d}v$ where $E(v)$ is the binding 
energy (per unit mass) of the system, and ${\cal F}(v)$ is the GW luminosity. 
$E(v)$ and ${\cal F}(v)$ are computed as post-Newtonian expansions in terms of 
$v$~\cite{Blanchet:LivRev}. Currently, the binding energy function $E(v)$ has 
been calculated to $v^6$ (3PN) accuracy by a variety of methods~\cite{DJS2000,
DJSequiv,BF00,BFeom,ABF01,BDE04,itoh1,itoh2}. The flux function ${\cal F}(v),$ 
on the other hand, has been calculated to $v^7$ (3.5PN) accuracy \cite{BFIJ02,
BDEI04} up to now only by the  multipolar-post-Minkowskian method and matching 
to a post-Newtonian source~\cite{Blanchet:LivRev}.

The inspiralling phase is usually pushed up to the point where the adiabatic evolution of circular
orbits breaks down due to the lack of further stable circular orbits. In the test-mass limit, the
last (or innermost) stable circular orbit (ISCO) can be computed exactly (at 6$M$ in Schwarzschild
coordinates). For comparable-mass binaries, on the other hand, the ISCO cannot always arise
unambiguously from PN theories.  In adiabatic models, the maximum-binding-energy condition (referred
to as MECO, or the maximum binding energy circular orbit,~\cite{Blanchet:2002}) can be used in place
of the ISCO.  This condition is reached when the derivative of the orbital binding energy with 
respect to orbital frequency vanishes. As a consequence, in this paper, the waveforms are evolved in 
time up to MECO: $E'(v)=0$.
It may be noted that the ISCO and MECO may
not be physically meaningful beyond the test-mass limit, but they make
convenient cutoff criteria. The appropriate region of validity of PN
waveforms can only be determined by comparison with fully general
relativistic results, such as the numerical simulations that we discussed 
earlier. 

Given $E(v)$ and ${\cal F}(v)$, one can construct different, but 
equivalent in terms of accuracy, approximations to the phasing 
by choosing to retain the involved functions or to re-expand them.
Indeed, the different PN models which describe the GW signal from 
inspiralling binaries agree with each other in the early stages of 
inspiral; but start to deviate in the late inspiral. The classification 
and explicit form of various models is nicely summarized in~\cite{DIS01}.
In this paper we use PN waveforms obtained by numerically solving
Eqs. (\ref{eq:phasing1}), called the {\it TaylorT1} approximant, to 
construct the `hybrid waveforms' (see Section~\ref{sec:Matching}). 

\subsection{Introduction to matched filtering}
Since we can model the signal reasonably well, it is natural to employ
matched filtering (which is the optimal detection strategy for a
signal of known shape in the stationary Gaussian noise) to search for
the gravitational-wave signal.  Suppose the detector's data $x(t)$
contains noise $n(t)$, and possible signal $s(t)$, i.e., $x(t) = n(t)
+ s(t)$.  Assuming $n$ to be stationary Gaussian noise, it is
convenient to work in the Fourier domain, because the statistical
property of the noise is completely characterized by its power
spectral density $S_n(f)$, which is given by (here we use a {\it
single-sided} spectrum)
\begin{equation}
\langle\tilde{n}(f)\tilde{n}^*(f')\rangle = \frac1{2}S_n(f)\,\delta(f-f')\,,
\end{equation}
where $\tilde n(f)$ is the Fourier Transform of $n(t)$ 
\begin{equation}
\tilde{n}(f) \equiv \int_{-\infty}^{\infty}n(t) e^{-2\pi \rmi   ft}\;\dt\,,
\end{equation}
and $\langle\ldots\rangle$ denotes taking the expectation value.
Based on the detector noise spectrum, we introduce a Hermitian inner
product:
\begin{equation}
(g | h) \equiv 2 \int_0^{\infty} \frac{\tilde{g}^*(f)\tilde{h}(f) + 
\tilde{g}(f)\tilde{h}^*(f)}{S_n(f)}\;\df\,.
\end{equation}
For the data $x$ with known signal $s$, the optimal detection statistic is given by 
applying a template $h$ with the same shape as $s$, or $h = \alpha s$:
\begin{equation}
\rho_{\rm opt} \equiv ( x | h)\,.
\end{equation}
The detectability of the signal is then determined by the SNR of $\rho_{\rm opt}$, 
\begin{equation}
\frac{S}{N} =
\left. \frac{(s|h)}{\sqrt{\langle(h|n)(n|h)\rangle}}\right|_{h=\alpha s} =
(s|s)^{1/2}. 
\end{equation}
(Note that the SNR does not depend on the overall normalization of $h$.) In case the 
template $h$ is not exactly of the same shape as $s$, the SNR will be reduced to 
\begin{equation}
\frac{S}{N} = (s|s)^{1/2}\mathcal{M}\,,
\end{equation}
where $\mathcal{M}\le 1$ is the {\it match} of the template to the signal, given by
\begin{equation}
\mathcal{M}[s,h]\equiv \frac{(s|h)}{\sqrt{(s|s)\,(h|h)}} \equiv (\hat
s|\hat h)\,,
\end{equation}
and where a hat denotes a normalized waveform.  For more details, we
refer the reader to Ref.~\cite{Cutler:1994ys}.

\subsection{Template banks, effectualness and faithfulness}

We now consider the more realistic problem of attempting to detect a
family of waveforms $s(\vectheta)$, parametrized by a vector of
physical parameters $\vectheta \in \Theta$, using a family of
templates $h(\veclambda)$ parametrized by a vector of 
parameters $\veclambda\in\Lambda$.  We first introduce the concepts of
{\it physical template bank} and {\it phenomenological template bank}.
Roughly speaking, physical template banks are constructed from
well-motivated physical models (e.g., approximation up to a certain
order)~ \cite{Babak:2006ty}, while phenomenological banks are constructed in an ad-hoc
manner to mimic the desired physical signals with high accuracy.  For
physical banks, the vectors $\vectheta$ and $\veclambda$ consists of
the same set of {\it physical parameters}, while for phenomenological
banks, the vector $\veclambda$ usually contains {\it phenomenological
  parameters,} which can be larger or smaller in number than the
physical parameters.  Two phenomenological template
families~\cite{BCV, BCV2} are used currently in the search for BH
binaries in LIGO data~\cite{Abbott:2007xi,Abbott:2005kq}.  They each
represent a different motivation for introducing phenomenological
banks: (i) when we have uncertainty in the signal model, we can
produce a template bank with larger detection efficiency by
introducing extra (phenomenological) parameters (BCV1, \cite{BCV}) so
that $\textrm{dim}(\Lambda) > \textrm{dim}(\Theta)$; (ii) when the
true signal depends on too many parameters and is too difficult to
search over, it is sometimes possible to come up with a model with
fewer (phenomenological) parameters ($\textrm{dim}(\Lambda) <
\textrm{dim}(\Theta)$) and still high fitting factors (BCV2,
\cite{BCV2}).

The detection efficiency of a template bank towards a specific signal
$s(\vectheta)$ can be measured by the threshold SNR above which the
detection probability exceeds a certain minimum (usually 50$\%$),
while the false-alarm probability is kept below a certain maximum
(usually 1$\%$ for one-year data). The threshold value depends
(logarithmically, in the case of Gaussian noise) on the number of
statistically independent templates, and (inverse-proportionally) on
the {\it fitting factor} (FF)~\cite{Apostolatos:1995pj}:
\begin{equation}
{\rm FF}[h;\vectheta] \equiv \max_{\veclambda}\mathcal{M}[s(\vectheta),h(\veclambda)]
\equiv \mathcal{M}[s(\vectheta),h(\veclambda_{\rm max})]  \,.
\end{equation}
A bank with high FF is said to be \emph{effectual}~\cite{DIS98,DIJS03}. Typically, we
require that the total mismatch between the template and true signal
(including the effects of both the fitting factor and the discreteness
of the template bank) to not exceed $3\%$.  We shall see that this
requirement is easily met by our template bank.

\begin{figure*}
  \begin{center}

    \psfrag{Pmap}{$P$}
    \psfrag{PMap2D}{$P_{2\rm D}$}
    \psfrag{Pt}{$P(\Theta)$}
    \psfrag{Pint}{$P_{\rm int}(\Theta)$}
    \psfrag{BT}{$\vectheta$}
    \psfrag{BLM}{$\veclambda_{\rm max}$}
    \psfrag{BLMP}{$\veclambda_{\rm max^\prime}$}
    \psfrag{BLMInt}{$\veclambda_{\rm int}$}
    \psfrag{Th}{$\Theta$}
    \psfrag{(i)}{(i)}
    \psfrag{(ii)}{(ii)}
    \psfrag{(iii)}{(iii)}

    \includegraphics[width=17.5cm]{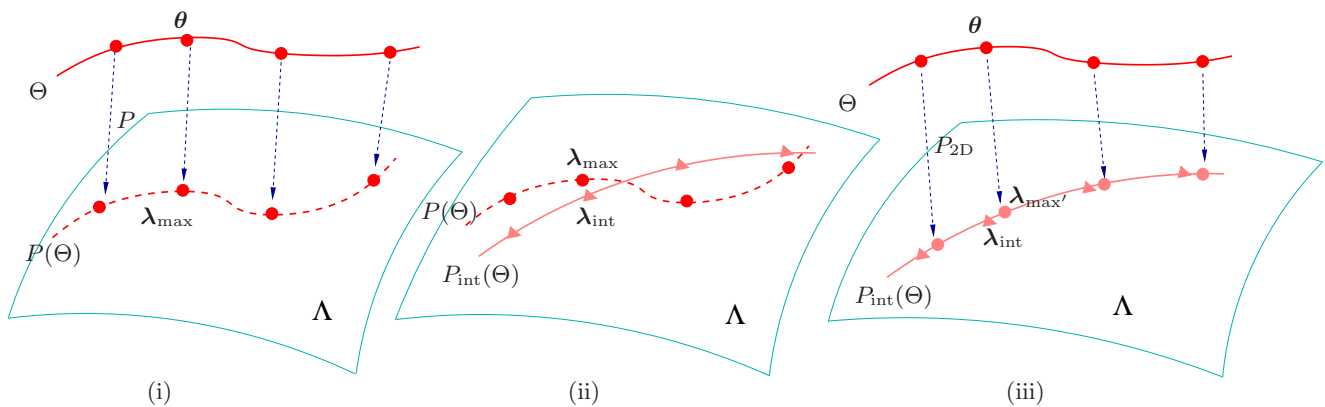}
    \caption{Construction of the phenomenological template bank: (i)
      mapping physical signals (solid curve) into a sub-manifold
      (dashed curve, with example templates marked by dots) of a
      larger-dimensional template bank (curved surface), (ii)
      obtaining a lower-dimensional phenomenological bank with the
      same number of parameters as physical parameters, through
      interpolation (solid curve on the curved surface, with example
      templates marked by triangles), and (iii) Estimating the bias of
      the lower-dimensional interpolated bank by mapping physical
      signals into the bank (with images of example signals marked by dots).}
    \label{fig:templSpaceCartoon}
  \end{center}
\end{figure*}

It is natural to associate every point $\vectheta$ in the physical
space $\Theta$ with the best matched point $\veclambda_{\rm
  max}\in\Lambda$. This leads to a mapping $P:\Theta \mapsto \Lambda$
defined by 
\begin{equation}
\label{def:P}
P (\vectheta) = \veclambda_{\rm max} \,.
\end{equation}
This mapping will play a key role in the construction of our template
bank.  We will assume the mapping $P$ to be single-valued, i.e., given
a target signal, the best-matched template is unique.  We depict this
mapping schematically in the left panel of
Fig.~\ref{fig:templSpaceCartoon}.

For a physical template bank with $\vectheta$ and $\veclambda$ the
same set of parameters (which we use $\vectheta$ to denote), it is
{\it most convenient} to identify the best-match parameter
$\vectheta_{\rm max}$ as the estimation of the original parameter
$\vectheta$. In general this will lead to a systematic bias 
\begin{equation}
\Delta\vectheta = \vectheta_{\rm max}-\vectheta = P(\vectheta)-\vectheta\,.
\end{equation}
A bank with a small bias (as defined above) is said to be \emph{faithful}~\cite{DIS98,DIJS03}.

However, if we assume no uncertainty in the true waveforms (thereby
excluding the case of BCV1), then as long as $P$ is invertible, a
non-faithful physical or phenomenological bank can always be {\it
converted into} a faithful bank by the re-parametrization
\begin{equation}
\label{faithful}
h_{\rm faithful}(\vectheta) \equiv h \circ P(\vectheta)
\end{equation}
where we have used the standard notation $h \circ P(\vectheta) :=h
(P(\vectheta))$.  In other words, each template $\veclambda$ in the
image set of physical signals $P(\Theta)$ is labeled by physical
parameters $\vectheta = P^{-1}(\veclambda)$.  For this reason, we
require $P$ to be invertible. It is quite conceivable that for
physical banks, $P$ should be invertible, if the physical bank
does not fail to describe the true waveforms too dramatically (and of
course assuming the true waveform does contain independent information
about the physical parameters $\vectheta$).  In this way, {\it all
  reasonable physical banks can be made faithful}.

By contrast, if for some phenomenological bank (e.g., BCV2 if we only
take into account the intrinsic parameters of the bank), $P$ is a
many-to-one map, with $P(\vectheta_1)=P(\vectheta_2)$ for some
$\vectheta_1 \neq \vectheta_2$. Then for a physical signal with
parameter $\vectheta_1$, the template bank $h_{\rm faithful}$ would
achieve the same best match at both $\vectheta_1$ and $\vectheta_2$,
making physical parameter determination non-unique. In this case, we
can simply keep using the phenomenological bank $h(\veclambda)$; once
a detection is made with $\veclambda_{\rm max}$, the a set of
parameters $P^{-1} (\veclambda_{\rm max})$ would be the best knowledge
we have about the physical parameters of the source. (In practice,
statistical uncertainty also applies to $\veclambda_{\rm max}$.)

\begin{figure*}
  \begin{center}
    \includegraphics[width=16cm]{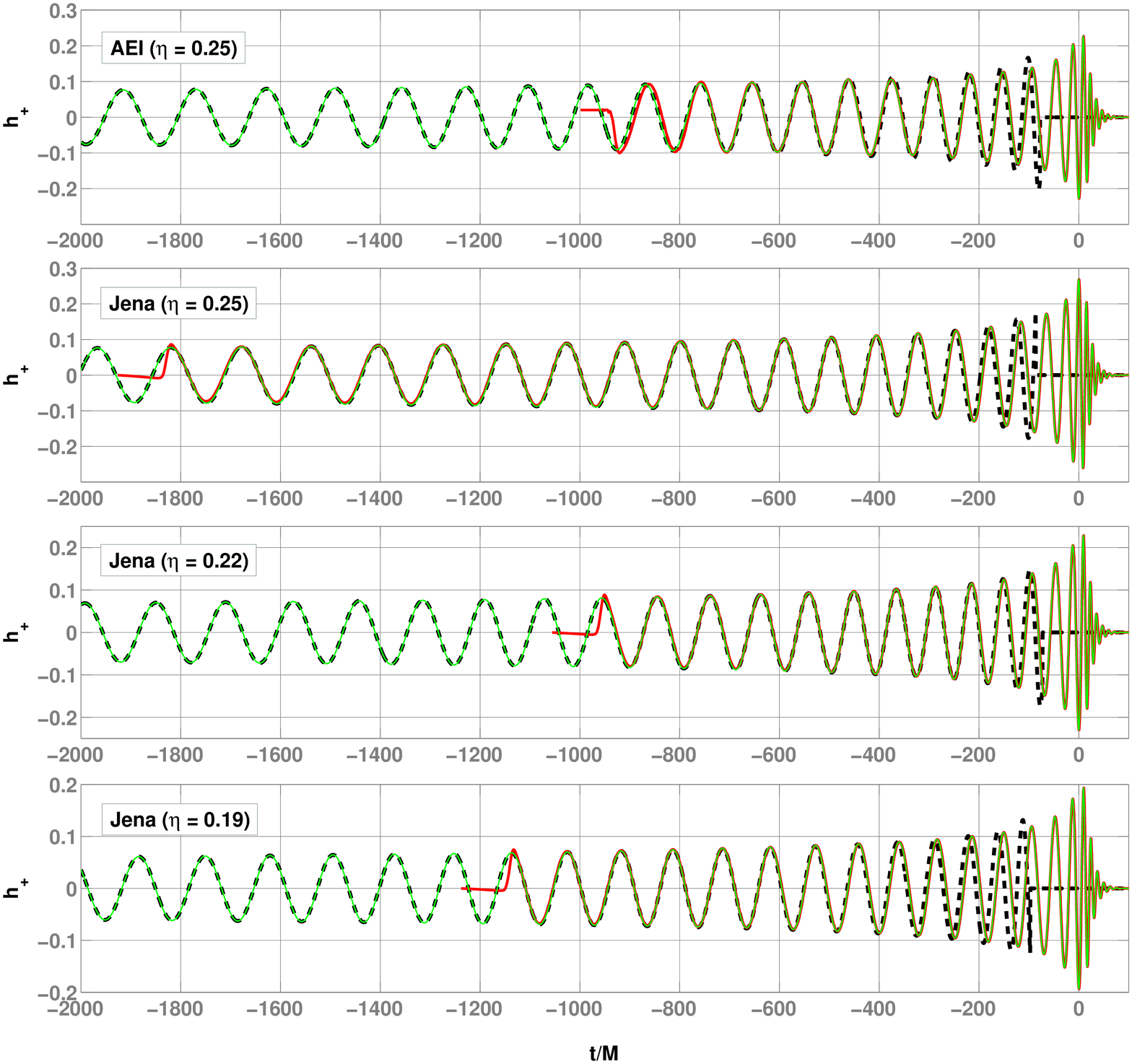}
    \caption{NR waveforms (thick/red), the `best-matched' 3.5PN
      waveforms (dashed/black), and the hybrid waveforms (thin/green)
      from three binary systems. The top panel corresponds to $\eta =
      0.25$ NR waveform produced by the AEI-CCT group. The second,
      third and fourth panels, respectively, correspond to $\eta =
      0.25, 0.22$ and $0.19$ NR waveforms from produced by the Jena
      group. In each case, the matching region is $-750 \leq t/M \leq
      -550$ and we plot the real part of the complex strain (the `+'
      polarization).}.
    \label{fig:timeDomWaveNRPNHyb}
  \end{center}
\end{figure*}

\section{A phenomenological template family for black-hole coalescence waveforms}
\label{sec:phenomtemplates}

\subsection{Strategy for constructing the phenomenological bank}
\label{sec:StrategyForTemplBank}

In our situation, since it is expensive to generate the entire
physical bank of templates using numerical simulations, we first
construct a highly effectual 10-dimensional phenomenological bank (motivated by the
format of PN waveforms), with effectualness confirmed by computing its
FF with a relatively small number of `target signals'.  Since we are
considering only non-spinning black holes, the physical parameter
space $\Theta$ is the set of all masses and symmetric mass ratios
$(M,\eta)$ that we wish to consider.  As we shall see shortly, for our
case the phenomenological parameter space $\Lambda$ is a
10-dimensional space.  Our templates will be denoted by
\begin{equation}
h(\veclambda) = h_{10\rm D}(\veclambda)\,.
\end{equation}
According to the discussion above
[Eqs.~\eqref{def:P}--\eqref{faithful}], if the mapping
$P:\Theta\mapsto\Lambda$ can indeed be obtained and inverted, then a
faithful two-dimensional (2D) phenomenological bank can be constructed
as
\begin{equation}
h_{2 \rm D}^{\rm faithful}(\vectheta) = h_{10\rm D} \circ P(\vectheta)\,.
\end{equation}

However, if our aim was to know $P$ exactly, then in principle we
would have to calculate accurate numerical waveforms for every
$(M,\eta)$ and to calculate the corresponding $\veclambda$ in each
case. This is obviously not practical, and we shall instead compute
$P$ at a few chosen points in $\Theta$ and interpolate to obtain an
approximation to $P$.  The detailed steps are as follows:
\begin{description}
\item[i.] While confirming effectualness of the ten-dimensional (10D)
  bank, we simultaneously obtain $N$ (a number manageable in terms of
  computational costs) data points for the mapping $P$,
  \begin{equation}
    \veclambda_{\rm max}^{(n)} = P(\vectheta^{(n)})\,,\quad n=1,2,\ldots,N
  \end{equation}
  which gives discrete points on the 2D manifold $P(\Theta)$. This is
  depicted by the left panel of Fig.~\ref{fig:templSpaceCartoon}.  
  
\item[ii.] Using these discrete points, we perform a smooth
  interpolation of $P$ denoted by $P_{\rm int}$. The form of $P_{\rm
    int}$ is motivated by PN waveforms, but with expansion
  coefficients determined by interpolation:
  \begin{equation}
    P_{\rm int} (\vectheta) = \veclambda_{\rm int}\,.
  \end{equation}
  This gives us a 2D phenomenological bank,
  \begin{equation}
    h_{2\rm D}(\vectheta) =h_{10\rm D}\circ P_{\rm int}(\vectheta)\,.
  \end{equation}
  This is depicted by the middle panel of
  Fig.~\ref{fig:templSpaceCartoon}. Due to the discrete choice of target
  waveforms, the constrained form of $P_{\rm int}$, and numerical errors
  (in the target waveforms as well as in searching for best-fit
  parameters), the interpolation will have errors, even at the sample
  points. This means the 2D bank will have slightly lower effectualness
  than the 10D bank.
  
\item[iii.] We re-test the effectualness of this 2D bank.  Note that
  there will be a new mapping $P_{2\rm D}$ which maps the physical
  parameters to the best fit parameters of this 2D bank.  We therefore find the
  best-matched parameters $\veclambda_{\rm max'}^{(n)}$, therefore
  obtaining discrete samples of the mapping $P_{\rm 2D}$:
  \begin{equation}
    \veclambda_{\rm max'}^{(n)} = P_{2\rm D} (\vectheta^{(n)})\,,
  \end{equation}
  yielding a systematic bias of
  \begin{equation}
    \Delta\vectheta^{(n)} = P_{\rm int}^{-1}(\veclambda_{\rm max'}^{(n)})
    - \vectheta^{(n)}\,. 
    \label{eq:Bias2DTemplBank}
  \end{equation}
  This is depicted in the right panel of Fig.~\ref{fig:templSpaceCartoon}.
\end{description}

In this paper, we construct the 2D template bank $h_{2\rm
D}(\vectheta)$ and estimate the systematic bias
$\Delta\vectheta^{(n)}$ in the estimation of parameters $\vectheta$,
as described above. But, it is also possible to construct an
interpolation $P_{2\rm D\,int}$ from the data points of $P_{2\rm D}$
so that we can construct a fully faithful (no systematic bias) bank
(up to interpolation error)
\begin{equation}
  h_{2\rm D}^{\rm faithful} (\vectheta) = h_{\rm 10 D}\circ P_{\rm int}
  \circ P_{\rm 2D\,int}(\vectheta). 
\end{equation}

\subsection{Constructing the `target signals'}
\label{sec:Matching}

The ultimate aim of this work is to create a family of
\textit{analytical} waveforms that are very close to the gravitational
waveforms produced by coalescing binary black holes. As a first step,
we need to construct a set of `target signals' containing all the
three (inspiral, merger and ring down) stages of the binary black hole
coalescence. Although numerical relativity, in principle, is able to
produce gravitational waveforms containing all these stages, the
numerical simulations are heavily constrained by their high
computational cost. It is therefore necessary, at the present time, to
use results from post-Newtonian theory to extend the waveforms
obtained from numerical relativity.

We produce a set of `hybrid waveforms' by matching the PN and NR waveforms in an 
overlapping time interval $t_1 \leq t < t_2$. The obvious assumption involved in this procedure 
is that such an overlapping region exists and that in it both approaches yield the
correct waveforms. These hybrid waveforms are assumed to be the target signals 
that we want to detect in the data of GW detectors. 

The NR and PN waveforms are given by Eq. (\ref{eq:NRh}) and Eq. (\ref{eq:PNhOptOrient}), 
respectively (with $\iota = 0$). The (complex) time-domain waveform $\hc(t,{\bm \mu})$ from a particular system
is parametrized 
by a set of `extrinsic parameters' ${\bm \mu} = \{\varphi_0, t_0\}$, where $\varphi_0$ 
is the initial phase and $t_0$ is the start time of the waveform.  We match the PN waveforms 
$\hc^{\rm PN}(t,{\bm \mu})$ and the NR waveforms $\hc^{\rm NR}(t,{\bm \mu})$
by minimizing the integrated squared absolute difference, $\delta$, between the two waveforms, \ie,
\ber
\delta &\equiv& \int_{t_1}^{t_2} \left|\hc^{^{\rm PN}}(t,{\bm \mu})-a\, \hc^{^{\rm NR}}(t,{\bm \mu})\right|^2 \, \dt.
\eer
The minimization is carried out over the extrinsic parameters ${\bm \mu}$ of the PN
waveform and an amplitude scaling factor $a$, while keeping the `intrinsic parameters'
($M$ and $\eta$) of both the PN and NR waveforms the same
\footnote{Here the amplitude scaling factor $a$ is introduced because of two reasons.
(i) The short NR waveforms used to construct the phenomenological template family
(see the following discussion in this Section) were extracted 
at a finite extraction radius. This introduces some error in the amplitude of the 
NR waveforms. (ii) Since the `long and accurate' NR waveforms (see the following discussion) are extrapolated to
an infinite extraction radius, we expect the amplitude of these waveforms to be
correct within numerical accuracy of the simulations. But, it turns out that the 
\emph{restricted} PN waveform has an amplitude which is inconsistent with the NR waveform
by roughly constant factor $6\pm 2\%$ in the frequency range we consider here \cite{Hannam:2007ik}.
For simplicity, we take the amplitude of the restricted PN waveform as the amplitude
scale for the hybrid waveforms. It should be noted that, since we use normalised 
templates, the errors that we introduce by this ($<10\%$) do not affect the fitting
factors or the detection statistic. But the horizon distance that we estimate in 
Section~\ref{sec:range} can have an error up to 10\% due to this choice.}. 
The hybrid waveforms are then produced by combining the `best-matched' PN
waveforms and the NR waveforms in the following way:
\be
\hc^{\rm hyb}(t, {\bm \mu}) \equiv a_0 \, \tau(t) \, \hc^{\rm NR}(t, {\bm \mu})+(1-\tau(t)) \, \hc^{\rm PN}(t, {\bm \mu_0})
\label{eq:HybWave}
\ee
where $\bm \mu_0$ and $a_0$ denote the values of $\bm \mu$ and $a$ for
which $\delta$ is minimized, and $\tau$ is a weighting function, defined as
\ber
\tau(t) \equiv \left\{ \begin{array}{ll}
0 & \textrm{if $t < t_1 $}\\ \\
\frac{t-t_1}{t_2-t_1}  & \textrm{if $t_1 \leq t < t_2 $}\\ \\
1 & \textrm{if $t_2 \leq t$.}
\end{array} \right.
\label{eq:HybWaveWeight}
\eer

In this paper we use two families of hybrid waveforms. Both are
produced by matching 3.5 PN TaylorT1 waveforms with NR waveforms. The
first set is constructed by using long ($> 10$ inspiral cycles) NR
waveforms. This include equal-mass ($\eta = 0.25$) NR waveforms
produced by the AEI-CCT group using their CCATIE code employing
fourth-order finite differencing to compute spatial derivatives, and
equal and unequal-mass ($\eta = 0.19, 0.22, 0.25$, or $M_1/M_2 =
1,2,3$) waveforms produced by the Jena group using their BAM code
employing sixth-order finite differencing and PN-motivated
initial-data parameters. The second set of hybrid waveforms is
constructed by using NR waveforms produced by the Jena group using
their BAM code employing fourth-order finite differencing. These are
short waveforms ($\sim 4$ inspiral cycles) densely covering a wide
parameter range ($0.16 \leq \eta \leq 0.25$).  We use the second set
of hybrid waveforms to construct the phenomenological family and to
test its efficiency in detecting signals from black hole coalescences,
and use the first set of hybrid waveforms (which are closer to the
actual signals) to verify our results.

The former family of hybrid waveforms is shown in Fig.~\ref{fig:timeDomWaveNRPNHyb}.  
The NR waveforms from three different simulations ($\eta = 0.25, 0.22, 0.19$) done 
by AEI and Jena groups are matched with 3.5PN inspiral waveforms 
over the matching region $-750 \leq t/M \leq -550$. The hybrid waveforms are
constructed by combining the above as per Eq. (\ref{eq:HybWave}) and Eq. (\ref{eq:HybWaveWeight}).

The robustness of the matching procedure can be tested by computing
the overlaps between hybrid waveforms constructed with different
matching regions. If the overlaps are very high, this can be taken as
an indication of the robustness of the matching procedure. A preliminary
illustration of this can be found in Ref.~\cite{Ajith:2007xh}, and a more
detailed discussion will be presented in~\cite{NRDApaper2}.

\begin{figure*}
  \begin{center}
    \includegraphics[height=7.5cm]{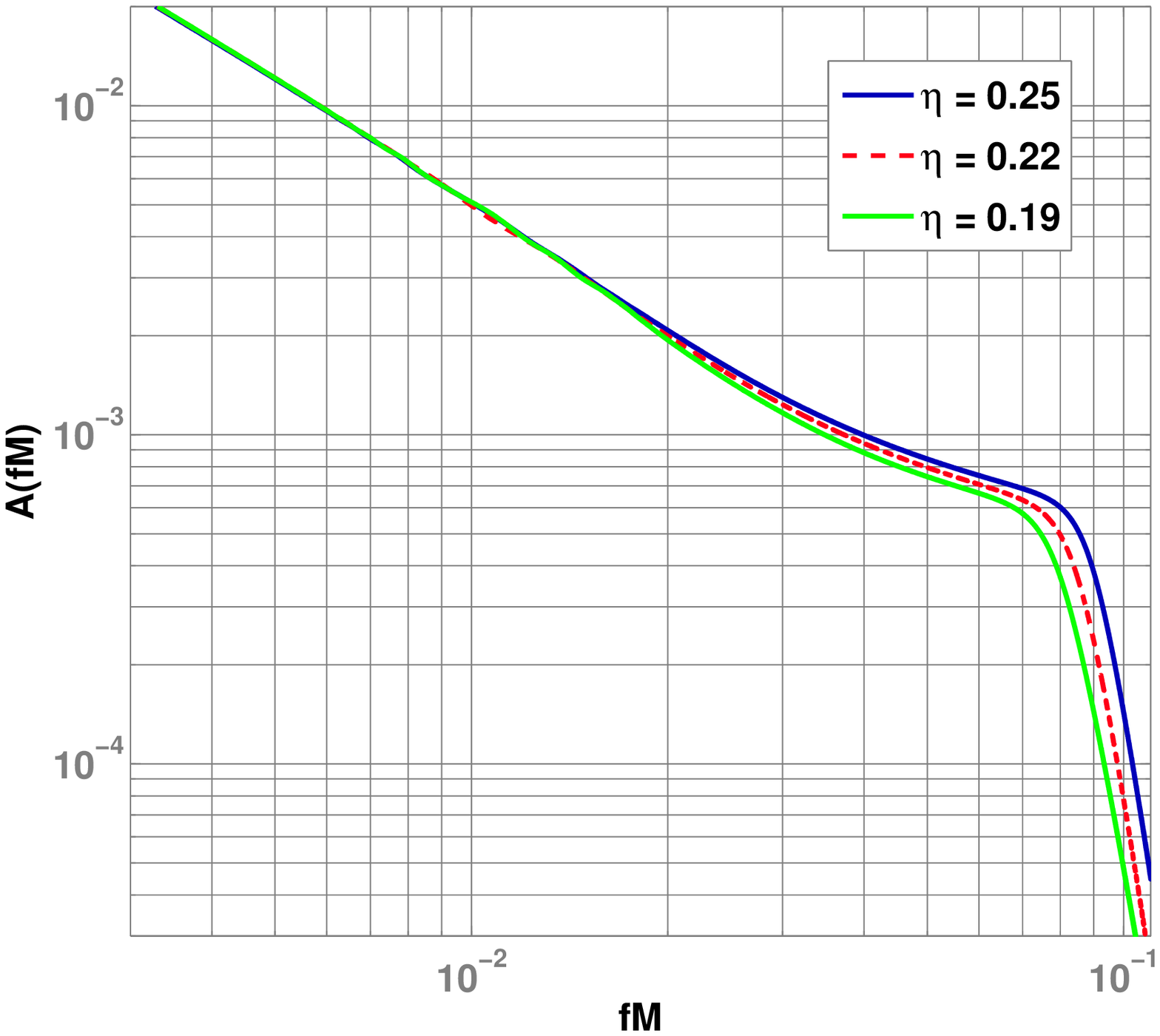}
    \includegraphics[height=7.5cm]{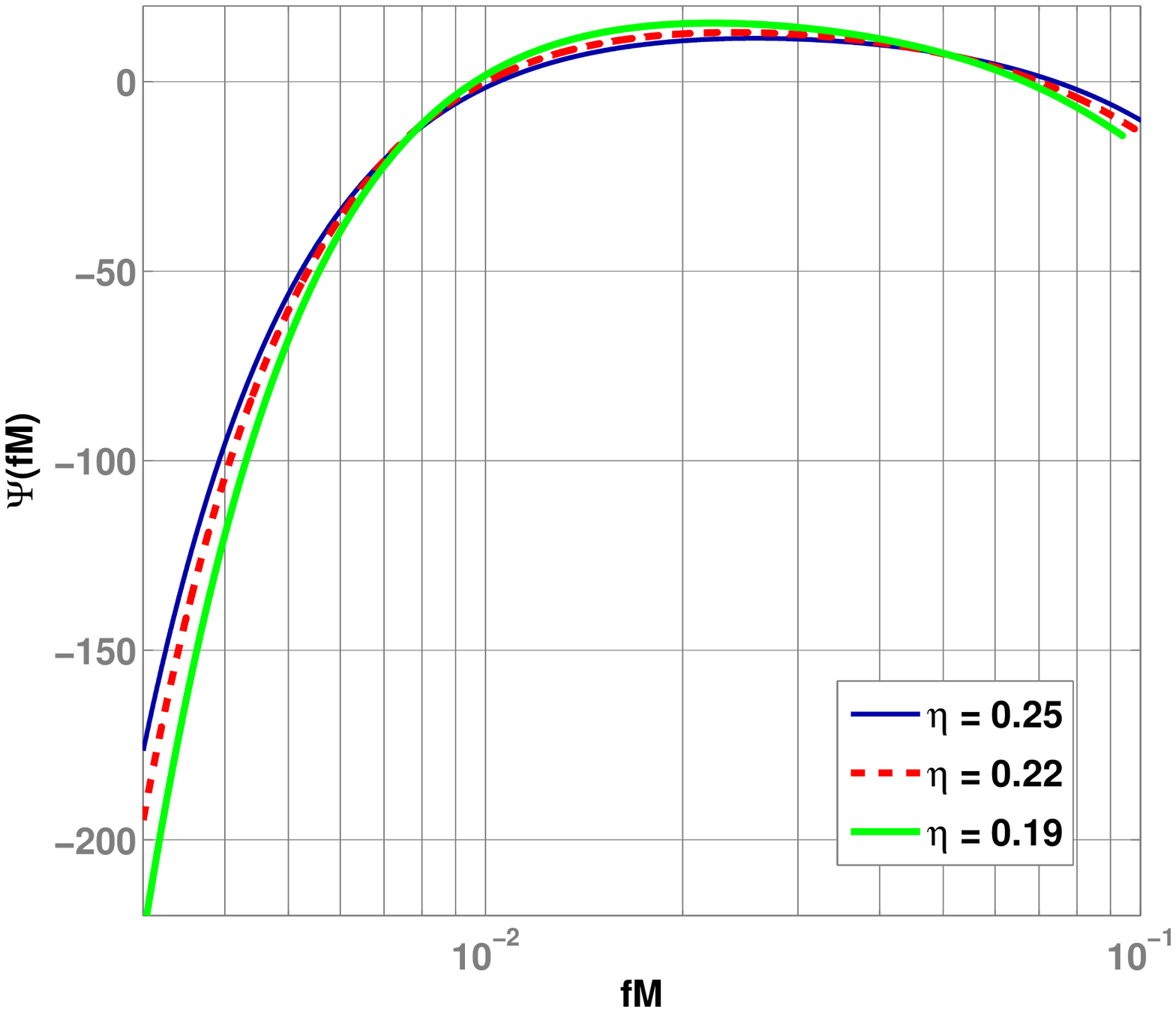}
    \caption{Fourier domain magnitude (left) and phase (right) of the
    (normalized) hybrid waveforms. 
    The constant phase term and the term linear in time (and frequency) have already been subtracted from the phase. 
    Symmetric mass-ratio $\eta$ of each waveform is
    shown in the legends. These waveforms are constructed by matching 3.5PN 
    waveforms with the long NR waveforms produced by the Jena group.}
    \label{fig:TargetWaveFreqDomain}
  \end{center}
\end{figure*}

Fig.~\ref{fig:TargetWaveFreqDomain} shows the hybrid waveforms of 
different mass-ratios in the Fourier domain. In particular, the panel 
on the left shows the amplitude of the waveforms in the Fourier domain, 
while the panel on the right shows the phase. These waveforms are constructed 
by matching 3.5PN waveforms with the long NR waveforms produced by the Jena
group. In the next section, we will try to parametrize these Fourier domain
waveforms in terms of a set of phenomenological parameters.

\subsection{Parametrizing the hybrid waveforms}
We propose a phenomenological parametrization to the hybrid waveforms in 
the Fourier domain. Template waveforms in the Fourier domain are of particular preference
because, (i) a search employing Fourier domain templates is computationally inexpensive 
compared to one using time domain templates (ii) parametrization of the hybrid 
waveforms is easier in the Fourier domain. 

We take our motivation from the restricted post-Newtonian approximation to model
the amplitude of the inspiral stage of the hybrid waveform, \ie, the amplitude is 
approximated to leading order as a power law $f^{-7/6}$ in terms of the Fourier frequency $f$
(as follows straight from adding leading order radiation reaction to Newtonian dynamics). The
amplitude of the merger stage is empirically approximated as a power law $f^{-2/3}$ 
(consistent with the observation of \cite{Buonanno:2006ui}), 
while the amplitude of the ring down stage is known to be a Lorentzian function 
around the quasi-normal mode ring down frequency. Similarly, we take our motivation 
from the stationary phase approximation (see, for example,~\cite{Thorne300}) of the inspiral waveform
to write the Fourier domain phase of the hybrid waveform as a series expansion in 
powers of $f$. As we shall see later, this provides an excellent approximation of the phase
of the hybrid waveform. 

\begin{figure}[tbh]
  \begin{center}
    \includegraphics[width=8.0cm]{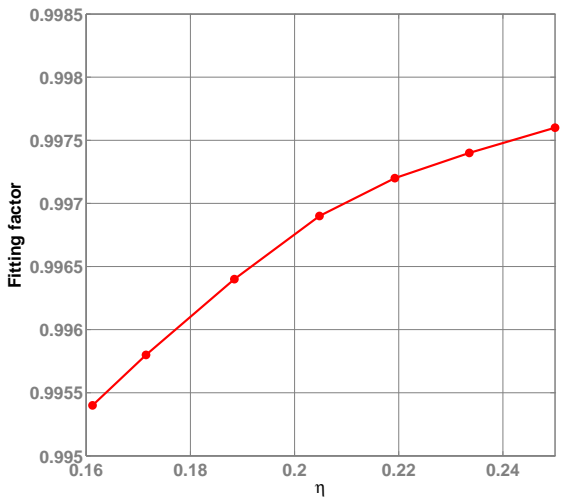}
    \caption{Fitting factors of the hybrid waveforms with the
    phenomenological waveform family.  Horizontal axis shows the
    symmetric mass ratio of the binary. Fitting factors are calculated 
    assuming a white noise spectrum, and hence are independent of the 
    mass of the binary.}
    \label{fig:FitFactorsJenaUM}
  \end{center}
\end{figure}

\begin{figure*}[t]
  \begin{center}
    \includegraphics[width=8cm]{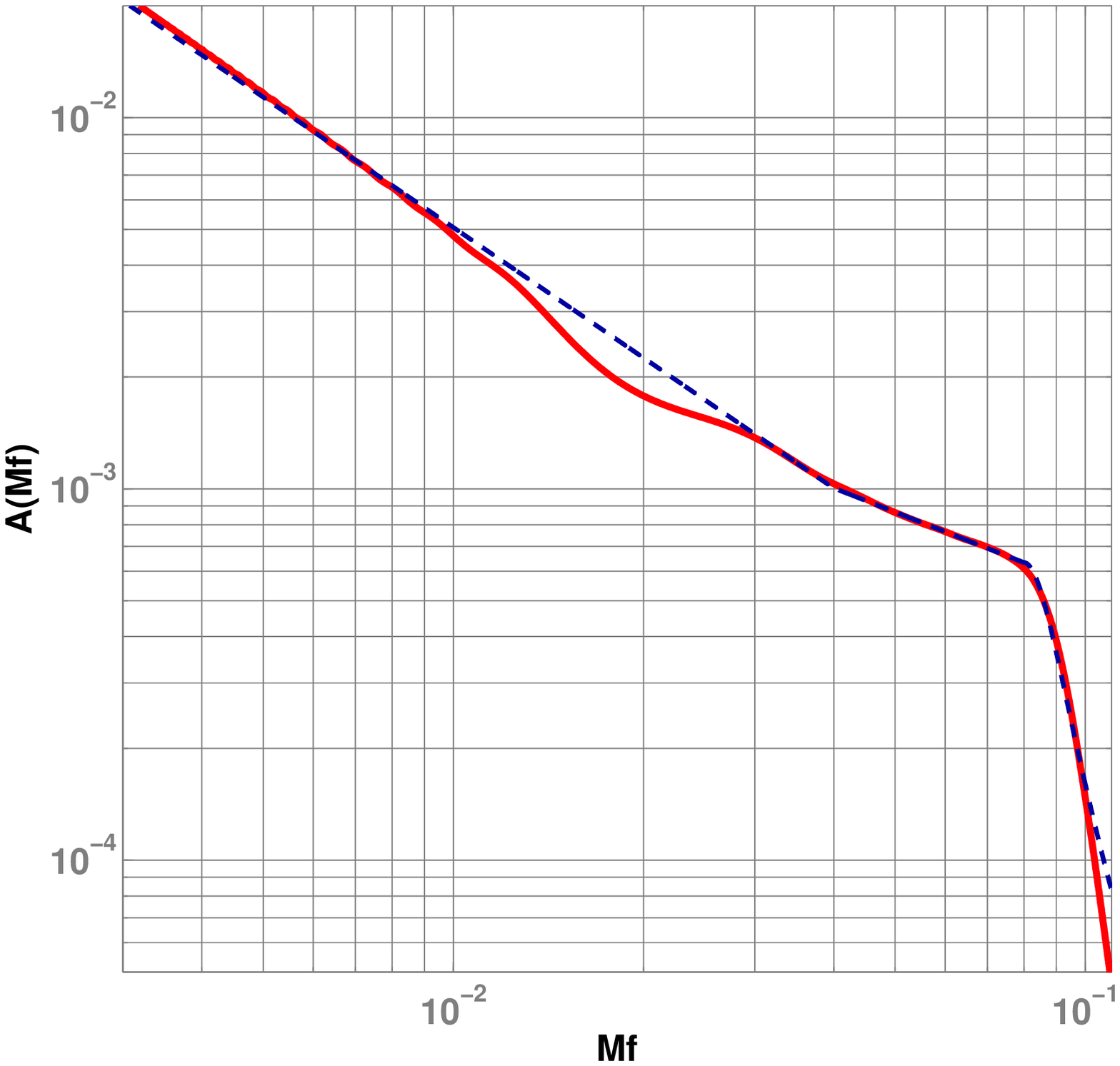}
    \includegraphics[width=8cm]{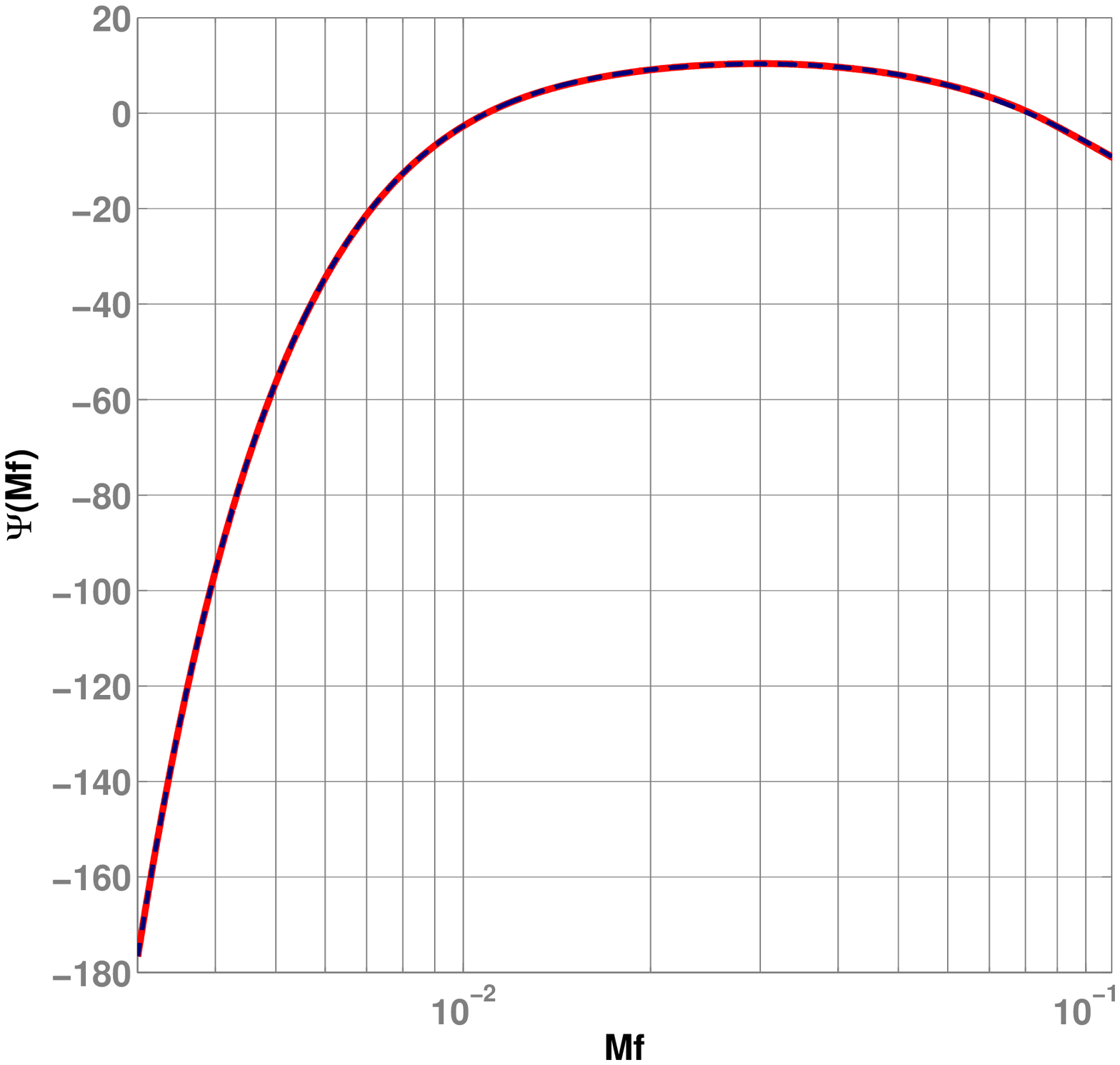}
    \caption{Hybrid waveforms (solid lines) in the frequency domain, and the `best-matched' phenomenological
    waveforms (dashed lines). The left panel shows the Fourier domain magnitude, while the right one
    shows the phase. 
    In the hybrid waveforms,  the constant phase term and the term linear in time (and frequency) have 
    already been subtracted from the phase.  In the phenomenological waveforms, $t_0$ and $\varphi_0$ 
    (see Eq.(\ref{eq:phenWavePhase})) have been chosen to minimize the phase difference between the hybrid
    and phenomenological waveforms. 
    These waveforms correspond to a binary with $\eta = 0.25$, and are constructed from the `short' NR
    waveforms produced by the Jena group (see Section~\ref{sec:Matching}). The `dip' in the 
    left panel at $Mf\simeq 2 \times 10^{-2}$ is due to the small eccentricity present in the 
    first few cycles of the NR waveform. All waveforms are normalized assuming a flat noise spectral density.}
    \label{fig:HybWaveAndBestMatchPhenWave}
  \end{center}
\end{figure*}

\subsubsection{Phenomenological waveforms}
We write our phenomenological waveform in the Fourier domain as 
\be
u(f) \equiv {A}_{\rm eff}(f) \, e^{{\mathrm i}\Psi_{\rm eff}(f)}.  
\label{eq:phenWave}
\ee 
where ${A}_{\rm eff}(f)$ is the amplitude of the waveform in the
frequency domain, which we choose to write in terms of a set of
\textit{`amplitude parameters'} $\bm \alpha = \{\fmerg, \fring, \sigma, \fcut\}$ as
\be
{A_{\rm eff}}(f) \equiv C
\left\{ \begin{array}{ll}
\left(f/\fmerg\right)^{-7/6} & \textrm{if $f < \fmerg$}\\\\
\left(f/\fmerg\right)^{-2/3} & \textrm{if $\fmerg \leq f < \fring$}\\\\
w \, {\cal L}(f,\fring,\sigma) & \textrm{if $\fring \leq f < \fcut$}
\end{array} \right. 
\label{eq:phenWaveAmp}
\ee
where $f_{\rm cut}$ is the cutoff frequency of the template and
$f_{\rm merg}$ is the frequency at which the power-law changes from
$f^{-7/6}$ to $f^{-2/3}$ (as noted previously in \cite{Buonanno:2006ui} for the equal-mass case).
$C$ is a numerical constant whose value depends on the relative orientations 
of the interferometer and the binary orbit as well as the physical parameters of 
the binary (see below). Also, in the above expression, 
\begin{equation}
{\cal L}(f,\fring,\sigma ) \equiv \left(\frac{1}{2 \pi}\right)
\frac{\sigma}{(f-\fring)^2+\sigma^2/4} \,,
\label{eq:Lorenzian}
\end{equation}
represents a Lorentzian function of width $\sigma$ centered around
$f_{\rm ring}$. The normalization constant $w$ is chosen in such a way
that ${A}_{\rm eff}(f)$ is continuous across the `transition'
frequency $f_{\rm ring}$, \ie,
\be
w \equiv \frac{\pi \sigma}{2} \left(\frac{f_{\rm ring}}{f_{\rm merg}}\right)^{-2/3}\,.
\ee

Taking our motivation from the stationary-phase approximation of the
gravitational-wave phase, we write the effective phase $\Psi_{\rm
eff}(f)$ as an expansion in powers of $f$,
\begin{equation}
 \Psi_{\rm eff}(f)  =  2 \pi f t_0 + \varphi_0 + \sum_{k=0}^{7} \psi_k\,f^{(k-5)/3}\,,
\label{eq:phenWavePhase}
\end{equation}
where $t_0$ is the time of arrival, $\varphi_0$ is the frequency-domain
phase offset, and $\bbeta = \{\psi_0,\,\psi_2,\,\psi_3,\,\psi_4,\,\psi_6,\,\psi_7\}$ are the
\textit{`phase parameters'}, that is the set of phenomenological
parameters describing the phase of the waveform.

The numerical constant $C$ in Eq.~(\ref{eq:phenWaveAmp}) can be determined by
comparing the amplitude of the phenomenological waveforms with that of the 
restricted post-Newtonian waveforms in the frequency domain. 

In the restricted post-Newtonian approximation, the Fourier transform of the gravitational
signal from an optimally-oriented binary located at an effective distance $d$ can be 
written as in Eq.~(\ref{eq:hOfFPNFreqDom}).  
We expect that in the inspiral stage ($f<\fmerg$) of our phenomenological waveforms
the amplitude will be equal to that of the post-Newtonian waveforms as given in 
Eq.~(\ref{eq:hOfFPNFreqDom}). Thus, in the case of an optimally-oriented binary, 
the numerical constant $C$ can be computed as
\be
C = \frac{M^{5/6}\,\fmerg^{-7/6}}{d\,\pi^{2/3}} \left( \frac{5\,\eta}{24} \right)^{1/2}.
\label{eq:phenWaveAmpConst}
\ee
This `physical' scaling will be useful when we estimate the sensitivity of a search using 
this template family (see Section~\ref{sec:range} and Appendix~\ref{app:horDist}).

We now compute the fitting factors of the hybrid waveforms with the family of 
phenomenological waveforms by maximizing the overlaps over all the parameters, 
\ie, $\{\balpha, \bbeta, \varphi_0, t_0\}$ of the phenomenological waveforms. 
While doing this, we also find the parameters, $\balpha_{\max}$ and $\bbeta_{\max}$, 
of the `best-matched' phenomenological waveforms. This calculation is described in 
detail in Appendix~\ref{app:fitfactor}. 

We first take a few (seven) hybrid waveforms coarsely spaced in the parameter range 
$0.16 \leq \eta \leq 0.25$, and compute the fitting factors and the best-matched 
phenomenological parameters, assuming a white-noise spectrum for the detector noise. 
We use these samples in the parameter space to construct the interpolated template 
bank (see next subsection). We then test the effectualness and faithfulness of the 
template bank using all ($\sim$ 30) hybrid waveforms finely spaced in the parameter space. 

The fitting factors are shown in Fig.~\ref{fig:FitFactorsJenaUM}. It is quite apparent that the 
fitting factors are always greater than 0.99, thus underlining the effectiveness of
the phenomenological waveforms in reproducing the hybrid ones. Also, as an example, in 
Fig.~\ref{fig:HybWaveAndBestMatchPhenWave}, we plot the hybrid waveforms from $\eta=0.25$
binary in Fourier domain along with the `best-matched' phenomenological waveform. 

\subsubsection{From phenomenological to physical parameters}

\begin{figure}[tb]
  \begin{center}
    \includegraphics[width=8.7cm]{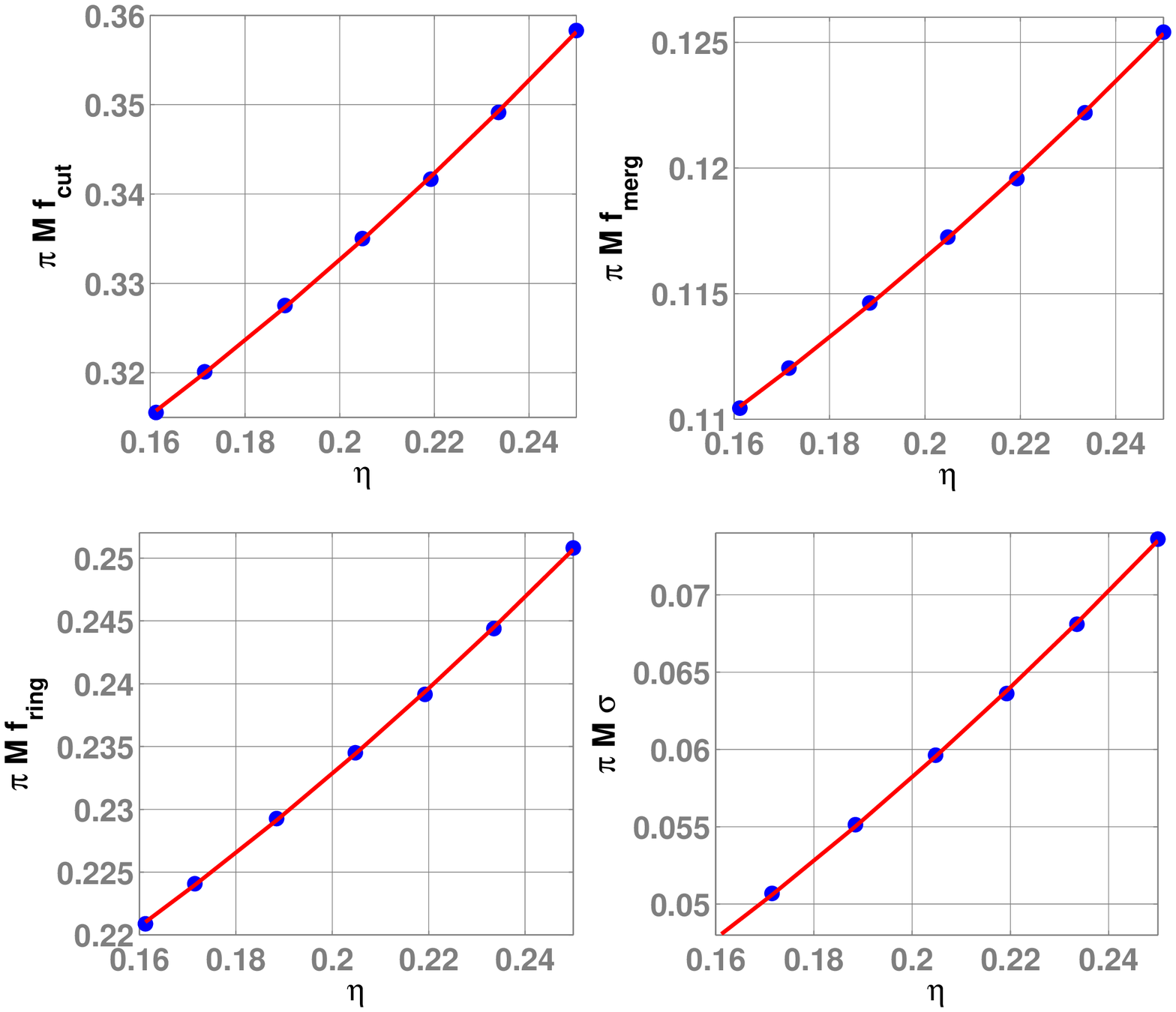}
    \caption{Best-matched amplitude parameters ${\balpha}_{\rm max}$
    in terms of the physical parameters of the binary (assuming white noise spectrum). The 
    horizontal axis shows the symmetric mass-ratio of the binary. Quadratic polynomial 
    fits ${\balpha}_{\rm int}$ to the data points are also shown.}
    \label{fig:BestMatchedAmpParams}
  \end{center}
\end{figure}

\begin{figure*}[t]
  \begin{center}
    \includegraphics[width=14.5cm]{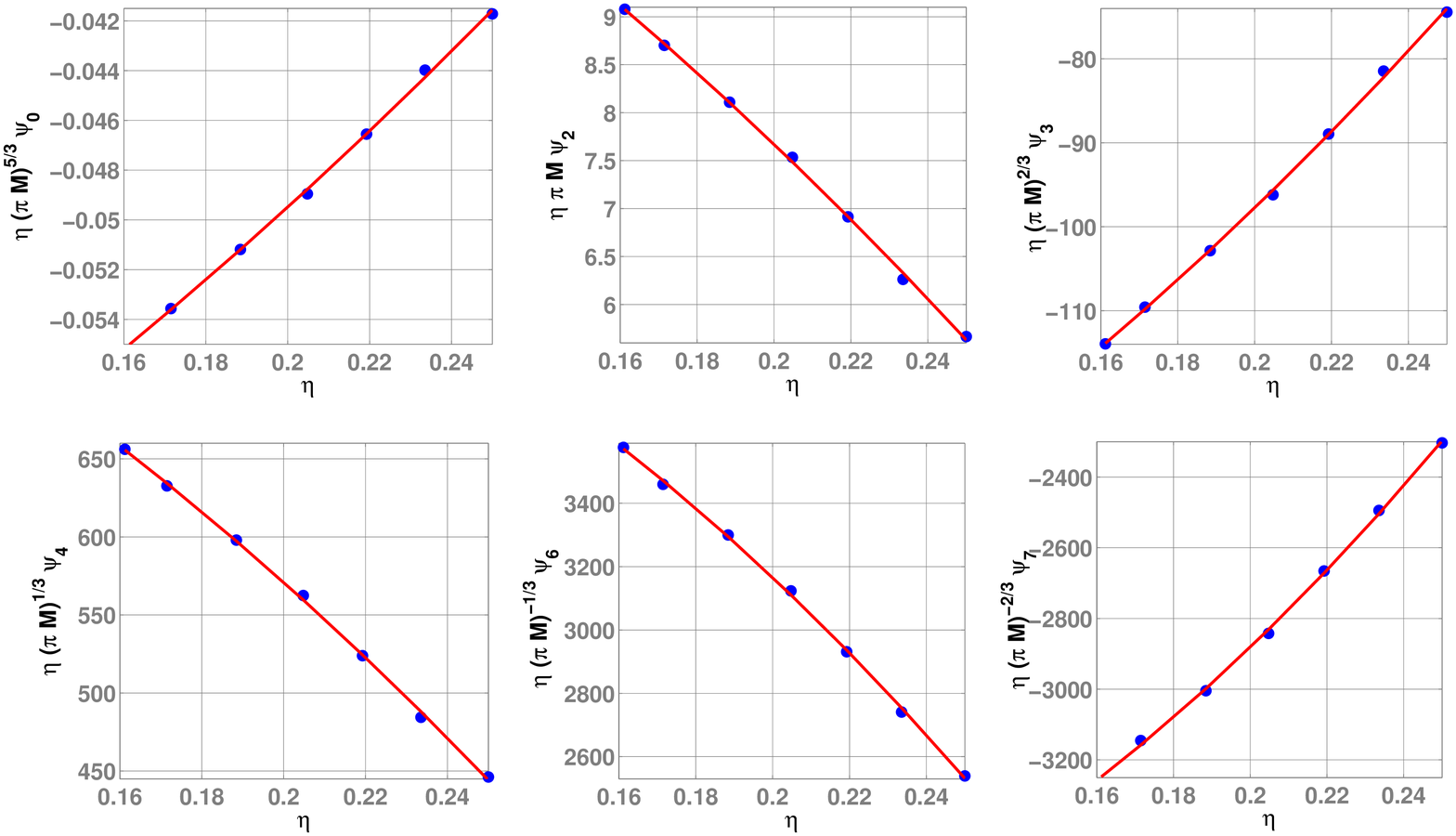}
    \caption{Best-matched phase parameters ${\bbeta}_{\rm max}$
    in terms of the physical parameters of the binary (assuming white noise spectrum). 
    The horizontal axis shows the symmetric mass-ratio of the binary. Quadratic polynomial 
    fits ${\bbeta}_{\rm int}$ to the data points are also shown.}
    \label{fig:BestMatchedPhaseParams}
  \end{center}
\end{figure*}

It is possible to parametrize the phenomenological waveforms having the 
largest overlaps with the hybrid waveforms in terms of the physical parameters 
of the hybrid waveforms. In Fig.~\ref{fig:BestMatchedAmpParams}, we plot the
amplitude parameters $\balpha_{\rm max}$ of the best-matched
phenomenological waveforms against the physical parameters of the
binary.  Similarly, the phase parameters ${\bbeta}_{\rm max}$ of
the best-matched phenomenological waveforms are plotted against the
physical parameters of the binary in Fig.~\ref{fig:BestMatchedPhaseParams}. 

It can be seen that $\balpha_{\rm max}$ and $\bbeta_{\rm max}$ can be 
written as quadratic polynomials in terms of the physical parameters ($M$ and $\eta$) of 
the hybrid waveforms as:
\ber
\alpha_{j~\rm int} &=& \frac{a_j \, \eta^2 + b_j \, \eta + c_j}{\pi M}  \,, \nonumber \\
\psi_{k~\rm int}  &=& \frac{x_k \, \eta^2 + y_k \, \eta + z_k}{ \eta \, (\pi M)^{(5-k)/3}}\,,
\label{eq:reparametr}
\eer
where $a_j,\, b_j,\, c_j,~j=0...3$ and $x_k,\, y_k,\, z_k,~k=0,\,2,\,3,\,4,\,6,\,7$ 
are the coefficients of the quadratic polynomials used to fit the data given
in Figs.~\ref{fig:BestMatchedAmpParams} and \ref{fig:BestMatchedPhaseParams}. 
These coefficients are listed in Tables~\ref{tab:polCoeffsAmpParams} and 
\ref{tab:polCoeffsPhaseParams}. It may be noted at this point that Figs.~\ref{fig:BestMatchedAmpParams} 
and \ref{fig:BestMatchedPhaseParams} correspond to the mapping $P:\vectheta^{(n)}\rightarrow \veclambda_{\max}^{(n)}$
that we have introduced in Section~\ref{sec:StrategyForTemplBank}, and Eq. (\ref{eq:reparametr}) to the 
interpolation $P_{\rm int}$ of $P$.


Using the empirical relations given in Eq. (\ref{eq:reparametr}), we can rewrite the effective 
amplitude and phase of the waveforms in terms of $M$ and $\eta$ as:

\begin{widetext}
\ber
{A_{\rm eff}}(f) &\equiv& C\,
\left\{ \begin{array}{ll}
\left(\frac{\pi M f}{a_0 \eta^2 + b_0 \eta + c_0}\right)^{-7/6} 
& \textrm{if~~$f < \frac{a_0 \eta^2 + b_0 \eta + c_0}{\pi M}$} \nonumber \\  \nonumber \\
\left(\frac{\pi M f}{a_0 \eta^2 + b_0 \eta + c_0}\right)^{-2/3} 
& \textrm{if~~$\frac{a_0 \eta^2 + b_0 \eta + c_0}{\pi M} \leq f < \frac{a_1 \eta^2 + b_1 \eta + c_1}{\pi M}$} \nonumber \\  \nonumber \\
w \, {\cal L}\left(f,\,\frac{a_1 \eta^2 + b_1 \eta + c_1}{\pi M},\,\frac{a_2 \eta^2 + b_2 \eta + c_2}{\pi M}\right) 
&\textrm{if~~$\frac{a_1 \eta^2 + b_1 \eta + c_1}{\pi M} \leq  f < \frac{a_3 \eta^2 + b_3 \eta + c_3}{\pi M}$}\,, \nonumber \\
\end{array} \right.\\  \nonumber \\
\Psi_{\rm eff}(f)  &=&  2 \pi f t_0 + \varphi_0 + \frac{1}{\eta}\,\sum_{k=0}^{7} (x_k\,\eta^2 + y_k\,\eta + z_k) \,(\pi M f)^{(k-5)/3}\,,
\eer
\end{widetext}
where the constant $C$ is given by Eq.(\ref{eq:phenWaveAmpConst}).
We use this family of parametrized waveforms to create a two-dimensional template bank of
non-spinning waveforms. This template family can be seen as a two-dimensional sub-manifold 
(parametrized by $M$ and $\eta$) embedded in a higher dimensional manifold (of the 
phenomenological waveforms).

The polynomial coefficients in the Table~\ref{tab:polCoeffsPhaseParams} are indeed significantly 
different from those predicted by stationary phase approximation of the PN inspiral phase in the 
frequency domain. There are two reasons for that: The first one is that our re-parametrization is 
optimized for the mass range where all three phases (inspiral, merger and ring down) are contributing 
significantly. The second reason is the residual eccentricity present in the numerical waveforms. Change 
in the relative significance of different PN terms reflects attempt to match the slightly eccentric 
waveform with circular.  When more accurate (less eccentric) numerical waveforms become available
in future, the re-parametrization given in Eq.(\ref{eq:reparametr}) can be optimized for a wider
mass range. An example of this can be seen in Ref.~\cite{Ajith:2007xh}. 

\subsection{Effectualness and faithfulness}

In order to measure the accuracy of our parametrized templates we
compute their overlap with the `target signals' (the hybrid
waveform).  To check the faithfulness of our phenomenological
templates, we compute their overlap with the target signal maximizing
it over the extrinsic parameters (time-of-arrival and the initial
phase).  We assess the effectualness of the parametrized waveforms by
computing fitting factors with the target signals (computing the
overlap maximized over both extrinsic and intrinsic
parameters). Faithfulness is a measure of how good the template
waveform is in both detecting a signal and estimating its
parameters. However, effectualness is aimed at finding whether or not
an approximate template model is good enough in detecting a signal
without reference to its use in estimating the parameters.

We compute the effectualness and the faithfulness of the template family for three different
noise spectra. The one-sided noise power spectral density (PSD) of the Initial LIGO detector
is given in terms of a dimensionless frequency $x=f/f_0$ by~\cite{LAL}
\begin{widetext}
\begin{equation}
S_h(f) =  9 \times 10^{-46} \left [ (4.49x)^{-56} + 0.16 x^{-4.52} + 0.52 + 0.32 x^2\right ]\,,
\end{equation}
where $f_0=150$ Hz; while the same for Virgo reads~\cite{LAL}
\begin{equation}
S_h(f) = 10.2 \times 10^{-46} \left [ (7.87 x)^{-4.8} + 6/17 x^{-1} + 1 + x^ 2 \right]\,,
\end{equation}
where $f_0=500$ Hz. For Advanced LIGO~\cite{LAL},
\begin{equation}
S_h(f) = 10^{-49} \left [x^{-4.14} - 5 x^{-2} + 111 \Big(\frac{1 - x^2 + x^4/2}{1 + x^2/2}\Big)\right]\,,
\end{equation}
\end{widetext}
where $f_0=215$ Hz. 

Faithfulness is computed by maximizing the overlaps over the extrinsic parameters $t_0$ and $\varphi_0$
only, which can be done trivially~\cite{schutz-91}. Effectualness is computed by maximizing
both intrinsic and extrinsic parameters of the binary. The maximization over the intrinsic parameters
is performed with the aid of the Nelder-Mead downhill simplex algorithm~\cite{amoeba}.  

\begin{figure*}
  \begin{center}
    \includegraphics[width=18cm]{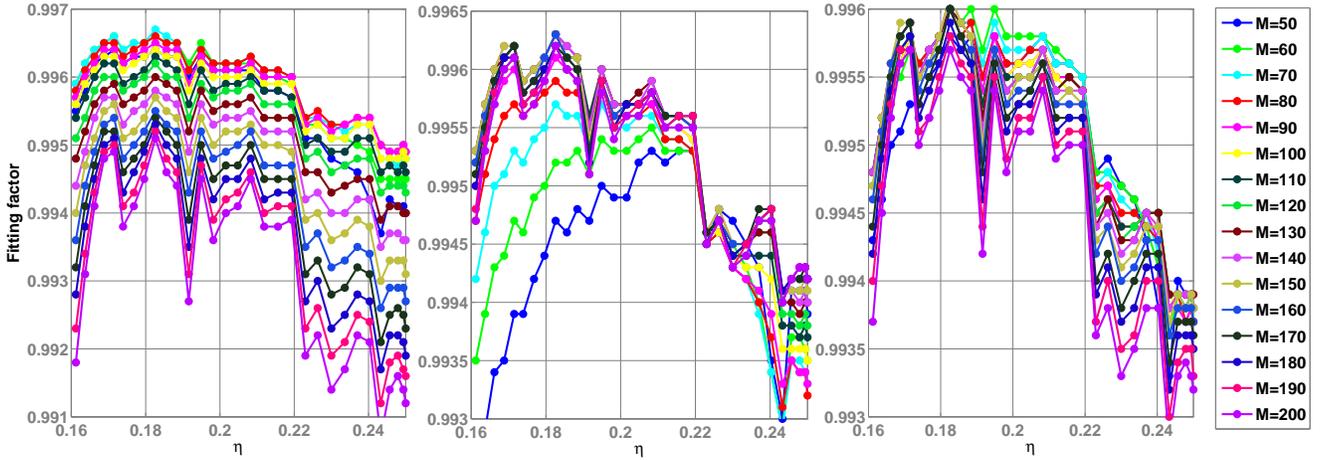}
    \caption{Fitting factor of the (two-dimensional) template family computed using three
    different noise spectra. The panel in the left correspond to the Initial LIGO noise PSD, 
    the one in the middle to the Virgo noise PSD and the one in the right to the
    Advanced LIGO noise PSD. 
    The horizontal axis represents the symmetric mass ratio $\eta$ of the binary and 
    the legends display the total mass $M$ (in units of $M_\odot$).}
    \label{fig:Effectualness}
  \end{center}
\end{figure*}

\begin{figure*}
  \begin{center}
    \includegraphics[width=18cm]{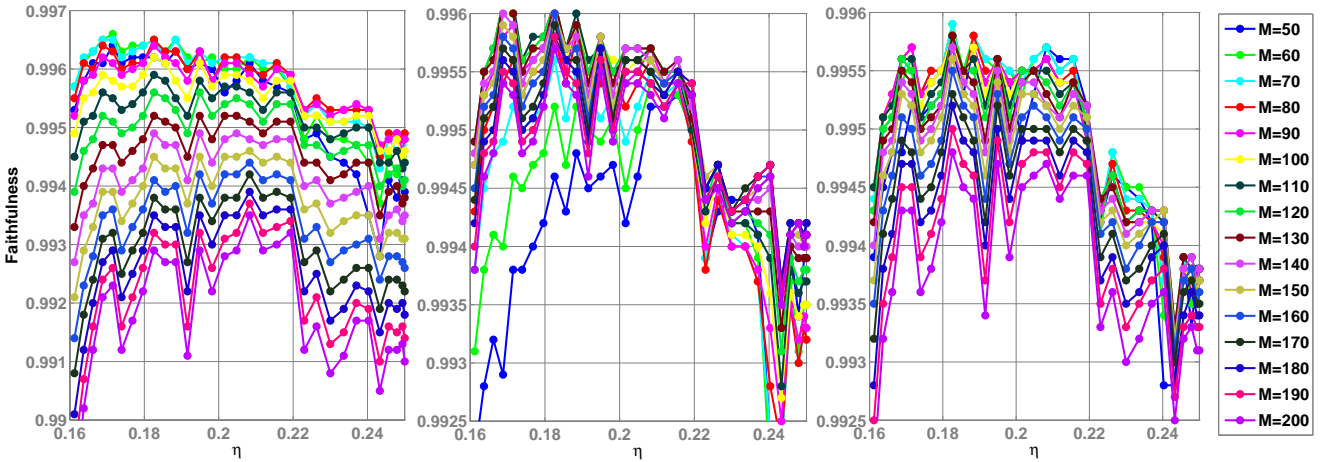}
    \caption{Same as in Fig.~\ref{fig:Effectualness}, except that the plots show 
    the faithfulness of the (two-dimensional) template family.}
    \label{fig:Faithfulness}
  \end{center}
\end{figure*}

The effectualness of the template waveforms with the hybrid waveforms is plotted in 
Fig.~\ref{fig:Effectualness} for three different noise spectral densities. The corresponding 
faithfulness is plotted in Fig.~\ref{fig:Faithfulness}. It is evident that, having both 
values always greater than 0.99, the proposed template family is both effectual and faithful.   

We also calculate the systematic bias in the estimation of parameters while maximizing 
the overlaps over the intrinsic parameters of the binary. The bias in the estimation 
of the parameters $\vectheta$ is defined in Eq.(\ref{eq:Bias2DTemplBank}). 

\begin{figure*}
  \begin{center}
    \includegraphics[width=18cm]{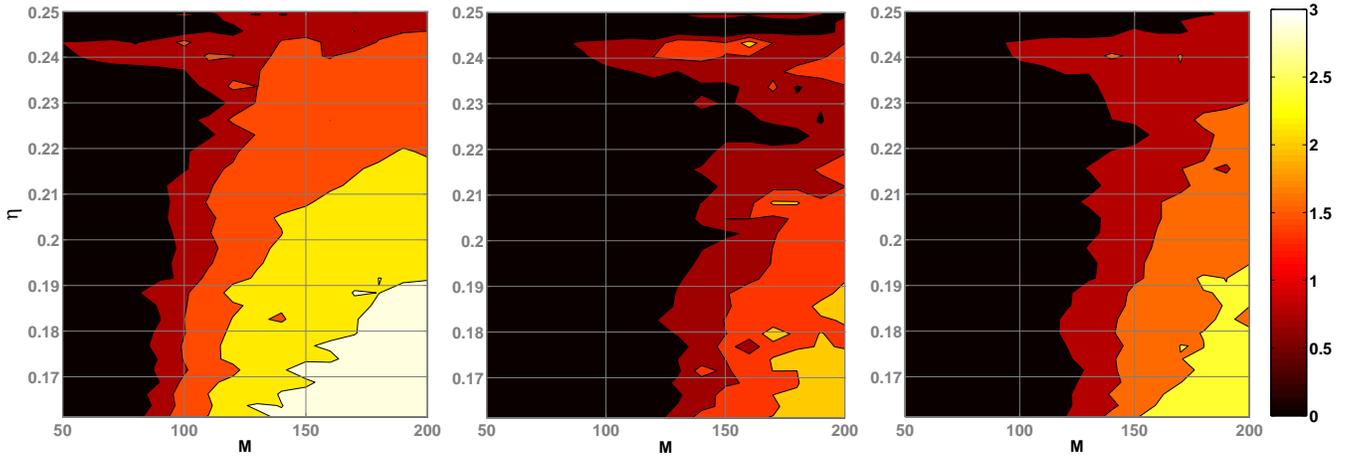}
    \caption{Bias in the estimation of $M$. Horizontal axis reports the total mass 
    $M$ (in units of $M_\odot$) and vertical axis reports the symmetric mass ratio 
    $\eta$ of the binary. Colors in the plot corresponds to the percentage bias, $|\Delta M|/M \times 100$. 
    The left panel corresponds to the Initial LIGO noise PSD, the middle panel to the Virgo noise 
    PSD and the right panel to the Advanced LIGO noise PSD.}
    \label{fig:biasPlotM}
  \end{center}
\end{figure*}

\begin{figure*}
  \begin{center}
    \includegraphics[width=18cm]{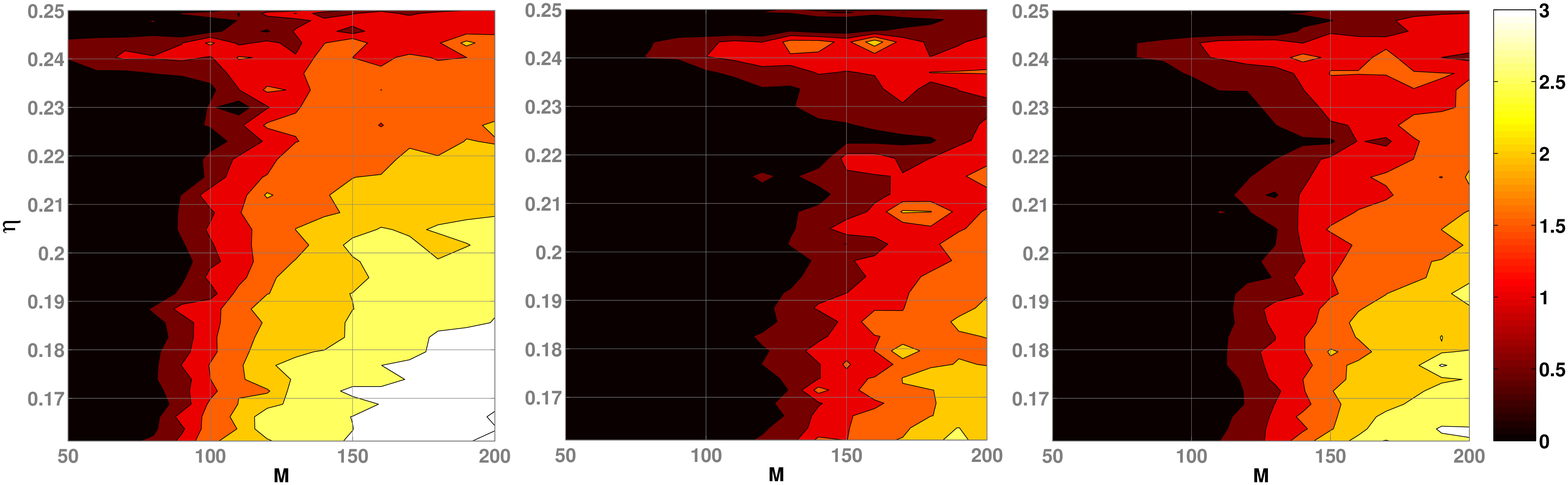}
    \caption{Same as in Fig.~\ref{fig:biasPlotM}, except that the plots show the percentage 
    bias $|\Delta \eta|/\eta \times 100$ in the estimation of $\eta$.}
    \label{fig:biasPlotEta}
  \end{center}
\end{figure*}

\begin{figure*}
  \begin{center}
    \includegraphics[width=18cm]{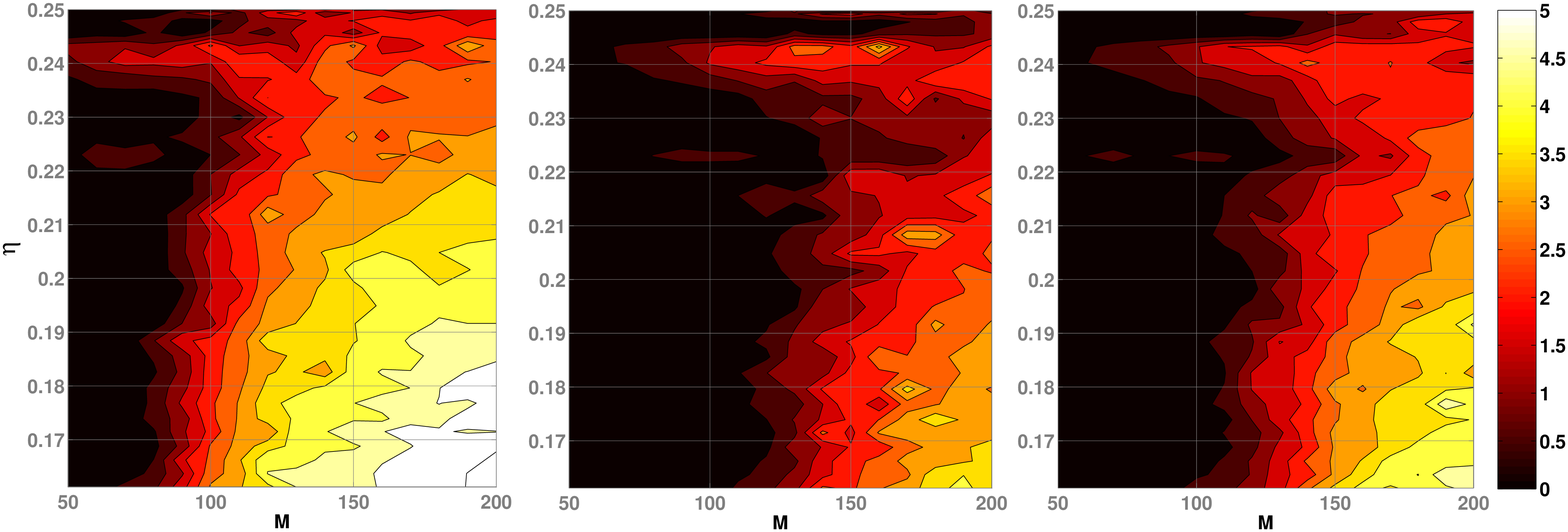}
    \caption{Same as in Fig.~\ref{fig:biasPlotM}, except that the plots show the percentage 
    bias $|\Delta \Mchirp|/\Mchirp \times 100$ in the estimation of $\Mchirp = M\eta^{3/5}$.}
    \label{fig:biasPlotMchirp}
  \end{center}
\end{figure*}

The percentage biases in estimating the total mass $M$, mass ratio $\eta$, and chirp mass 
$\Mchirp = M\eta^{3/5}$ of the binary are plotted in Figs.~\ref{fig:biasPlotM}, \ref{fig:biasPlotEta}, 
and \ref{fig:biasPlotMchirp}, respectively. This preliminary investigation suggests that the 
bias in the estimation of $M$ and $\eta$ using the proposed template family is $<3\%$, while the same in 
estimating $\Mchirp$ is $<6\%$. 

\begin{table}[tbh]
    \begin{center}
        \begin{tabular}{cccccccc}
            \hline
            \hline
            Parameter &\vline& \multicolumn{1}{c}{$a_k$} &\vline& \multicolumn{1}{c}{$b_k$} &\vline& \multicolumn{1}{c}{$c_k$} \\
            \hline
            $\fmerg$ &\vline& 2.9740$\times 10^{-1}$   &\vline& 4.4810$\times 10^{-2}$    &\vline&  9.5560$\times 10^{-2}$    \\
            $\fring$ &\vline& 5.9411$\times 10^{-1}$   &\vline& 8.9794$\times 10^{-2}$    &\vline& 1.9111$\times 10^{-1}$     \\
            $\sigma$ &\vline& 5.0801$\times 10^{-1}$   &\vline& 7.7515$\times 10^{-2}$    &\vline&  2.2369$\times 10^{-2}$    \\
            $\fcut$  &\vline& 8.4845$\times 10^{-1}$   &\vline& 1.2848$\times 10^{-1}$    &\vline&  2.7299$\times 10^{-1}$    \\
            \hline
            \hline
        \end{tabular}
        \caption{Polynomial coefficients of the best-matched amplitude parameters. The 
        first column lists the amplitude parameters $\balpha_{\rm int}$. Eq.(\ref{eq:reparametr}) shows 
        how these parameters are related to the coefficients $a_k, b_k, c_k$.}
        \label{tab:polCoeffsAmpParams}
    \end{center}
\end{table}

\begin{table}[tbh]
    \begin{center}
        \begin{tabular}{ccccccccc}
            \hline
            \hline
            Parameter &\vline& \multicolumn{1}{c}{$x_k$} &\vline& \multicolumn{1}{c}{$y_k$} &\vline& \multicolumn{1}{c}{$z_k$}  \\
            \hline
            $\psi_0$ &\vline& 1.7516$\times 10^{-1}$   &\vline& 7.9483$\times 10^{-2}$    &\vline& -7.2390$\times 10^{-2}$    \\
            $\psi_2$ &\vline& -5.1571$\times 10^{1}$   &\vline&-1.7595$\times 10^{1}$     &\vline& 1.3253$\times 10^{1}$      \\
            $\psi_3$ &\vline& 6.5866$\times 10^{2}$    &\vline& 1.7803$\times 10^{2}$     &\vline& -1.5972$\times 10^{2}$     \\
            $\psi_4$ &\vline& -3.9031$\times 10^{3}$   &\vline&-7.7493$\times 10^{2}$     &\vline& 8.8195$\times 10^{2}$      \\
            $\psi_6$ &\vline& -2.4874$\times 10^{4}$   &\vline&-1.4892$\times 10^{3}$     &\vline& 4.4588$\times 10^{3}$      \\
            $\psi_7$ &\vline& 2.5196$\times 10^{4}$    &\vline& 3.3970$\times 10^{2}$     &\vline& -3.9573$\times 10^{3}$     \\
            \hline
            \hline
        \end{tabular}
        \caption{Polynomial coefficients of the best-matched phase parameters.  The 
        first column lists the phase parameters $\bbeta_{\rm int}$. Eq.(\ref{eq:reparametr}) shows
        how these parameters are related to the coefficients $x_k, y_k, z_k$.}
        \label{tab:polCoeffsPhaseParams}
    \end{center}
\end{table}

\subsection{Verification of the results using more accurate hybrid waveforms}
\label{sec:Verification}

As we have discussed in Section~\ref{sec:Matching}, the hybrid waveforms used for constructing
the template waveforms are produced by matching rather short ($\sim$ 4 inspiral cycles)
NR waveforms with PN waveforms. We have also produced a few hybrid waveforms by matching PN waveforms
with long ($>$ 10 inspiral cycles) and highly accurate (sixth-order finite differencing and
low eccentricity) NR waveforms. This set of hybrid waveforms (which are closer
to the `actual signals') can be used to verify the efficacy of the template waveforms 
in reproducing these more accurate signals. 

\begin{figure*}
  \begin{center}
    \includegraphics[width=15cm]{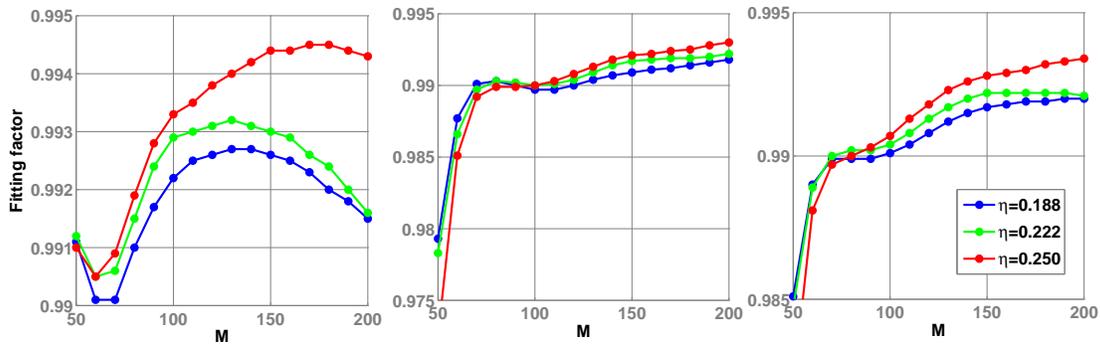}
    \caption{Fitting factor of the (two-dimensional) template family with `more accurate'
    hybrid waveforms (see Section~\ref{sec:Verification}). The overlaps are computed using three
    different noise spectra. The panel in the left correspond to the Initial LIGO PSD, 
    the one in the middle to the Virgo noise PSD and the one in the right to the 
    Advanced LIGO noise PSD. 
    The horizontal axis represents the total mass $M$ (in units of $M_\odot$) and 
    the legends display the symmetric mass ratio $\eta$ of the binary.}
    \label{fig:FitFactWithJenaLongUM}
  \end{center}
\end{figure*}

Fig.~\ref{fig:FitFactWithJenaLongUM} shows the fitting factors of the two-dimensional
template family with the `more accurate' hybrid waveforms. The fitting factors are 
computed, as before, using the Initial LIGO (left), Virgo (middle) and Advanced LIGO (right) 
noise spectra. The high fitting factors (although smaller than the same obtained 
in the previous Section) with the hybrid waveforms once again underline the 
efficacy of the template waveforms in reproducing the hybrid ones. It is indeed expected that the template family
will have better overlaps with the hybrid waveforms described in the 
previous Section (those constructed from `short' NR waveforms), because the polynomial coefficients
given in Tables~\ref{tab:polCoeffsAmpParams} and \ref{tab:polCoeffsPhaseParams} are 
optimized for these hybrid waveforms. When more `long and accurate' NR 
waveforms are available in the future, the polynomial coefficients given in the Tables
can be optimized for the corresponding family of `more accurate' hybrid waveforms. 
In any case, since the fitting factors are already very high, we don't expect any 
significant improvements. 

\subsection{The astrophysical range and comparison with other searches}
\label{sec:range}

\begin{figure*}
  \begin{center}
    \includegraphics[width=18cm]{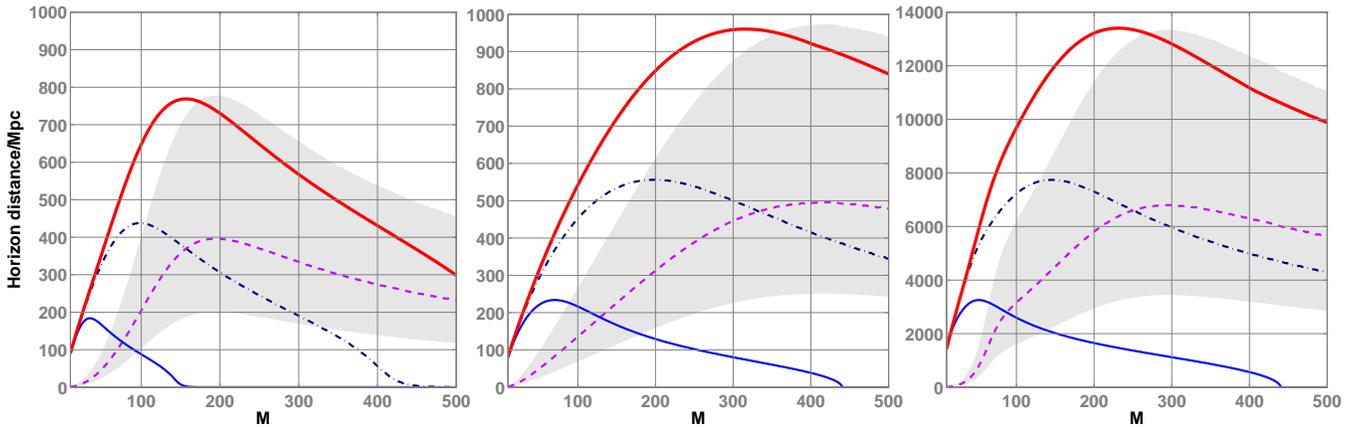}
    \caption{Distance to optimally located and oriented equal-mass
      binaries which can produce an optimal SNR of 8 at the Initial
      LIGO (left plot), Virgo (middle plot) and Advanced LIGO (right
      plot) noise spectra. Horizontal axis reports the total mass of
      the binary (in units of $M_\odot$) and vertical axis reports the
      distance in Mpc.  In each plot, the thin solid (blue) line
      corresponds to a search using standard PN templates truncated at
      ISCO, the dotted-dashed (black) line to a search using effective
      one body waveform templates truncated at the light ring, the the
      dashed (purple) line to a search using ring-down templates, and
      the thick solid (red) line to a search using the template family
      proposed in this paper. The ring down horizon distance is
      computed assuming that $\epsilon=0.7\%$ of the black hole mass
      is radiated in the ring down stage, while the Kerr parameter
      $a=0.69$ is known from the numerical simulation. Since the value
      of $\epsilon$ has some amount of uncertainty in it, we have also
      included the shaded region in the plot corresponding to $0.18\%
      \leq \epsilon \leq 2.7\%$.
    }
    \label{fig:HorDist}
  \end{center}
\end{figure*}

The template family proposed in this paper can be used for coherently
searching for all the three stages (inspiral, merger, and ring-down)
of the binary black hole coalescence, thus making this potentially more sensitive
than searches which look at the three stages separately.
Fig.~\ref{fig:HorDist} compares the sensitivity of the searches
using different template families. What is plotted here are the
distances at which an optimally-oriented, equal-mass binary would produce an
optimal SNR of 8 at the Initial LIGO (left plot), 
Virgo (middle plot) and Advanced LIGO (right plot) noise spectra. 
In each plot, the thin solid (blue) line corresponds to a search using PN templates
truncated at the innermost stable circular orbit (ISCO) of the Schwarzschild geometry 
having the same mass as the total mass $M$ of the binary; 
the dashed (purple) line to a search using ring-down templates~\cite{Goggin:2006bs}; 
the dot-dashed (black) line to a search using effective one body~\cite{Buonanno:1998gg} waveform
templates truncated at the light ring of the corresponding Schwarzschild geometry,  
and the solid line to a search using all
three stages of the binary coalescence using the template bank
proposed here. The computation is described in detail in Appendix~\ref{app:horDist}. 
The horizontal axis reports the total-mass of the
binary, while the vertical axis the distance in Mpc. It is quite
evident that, for a substantial range of total mass ($100 \lesssim M/M_\odot \lesssim 300$ for Initial LIGO, 
$200 \lesssim M/M_\odot \lesssim 500$ for Virgo, $150 \lesssim M/M_\odot \lesssim 400$ for Advanced LIGO), the
`coherent search' using the new template family is significantly more 
sensitive than any other search considered here. 
 
However, while this looks promising, we repeat here the caveats
emphasized in \cite{Ajith:2007qp}: It is important
to treat Fig.~\ref{fig:HorDist} as only a preliminary assessment;
fitting factors are not the only consideration for a practical search
strategy. It is also very important to consider issues which arise
when dealing with real data.  For example, false alarms produced by
noise artifacts might well determine the true sensitivity of the
search, and these artifacts will inevitably be present in real data.
This is however beyond the scope of the present work, and further
investigation is required before we can properly assess the efficacy
of our phenomenological template bank in real-life searches.

\section{Summary and outlook}
\label{sec:summary}

Making use of the recent results from numerical relativity we have
proposed a phenomenological waveform family which can model the
inspiral, merger and ring-down stages of the coalescence of
non-spinning binary black holes in quasi-circular orbits. We first
constructed a set of hybrid waveforms by matching the NR waveforms
with analytical PN waveforms. Then, we constructed analytical
phenomenological waveforms which approximated the hybrid waveforms.
The family of phenomenological waveforms that we propose was found to
have fitting factors larger than 0.99 with the hybrid waveforms.  We
have also shown how this phenomenological waveform family can be
parametrized solely in terms of the physical parameters ($M$ and
$\eta$) of the binary, so that the template bank is, in the end, two
dimensional~\footnote{It may be noted that, the mapping from the 
phenomenological to physical parameters might not be unique in the case 
of spinning binaries, because of the degeneracies of different spin configurations.}. 
This two dimensional template family can be explicitly
expressed in terms of the physical parameters of the binary. We have
estimated the `closeness' of this two-dimensional template family
with the family of hybrid waveforms in the detection band of three
ground-based GW detectors, namely Initial LIGO, Virgo and Advanced
LIGO. We have estimated the effectualness (larger overlaps with the
target signals for the purpose of detection) and faithfulness (smaller
biases in the estimation of the parameters of the target signals) of
the template family. Having both types of overlap always greater than
0.99, the two dimensional template family is found to be both
effectual and faithful in the detection band of these ground-based
detectors.

This phenomenological waveform family can be used to densely cover the 
parameter space, avoiding the computational cost of generating numerical 
waveforms at every grid point in the parameter space. We have compared 
the sensitivity of a search using this template family with other searches. 
For a substantial mass-range, the search using all three stages of the 
binary black hole coalescence was found to be significantly more sensitive
than any other template-based searches considered in this paper. This might enable 
us to do a more sensitive search for intermediate-mass black holes using 
ground-based GW detectors.

A number of practical issues need to be addressed before we can employ this 
template family in an actual search for GW signatures. The first issue will be how to
construct a bank of templates sufficiently densely spaced in the parameter space so 
that the loss in the event rate because of the mismatch between the signal and template
is restricted to an acceptable amount (say, 10\%). The explicit frequency domain 
parametrization of the proposed template family makes it easier to adopt the formalism
proposed by Owen~\cite{Owen:1995tm} in laying down the templates using a metric in the 
parameter space. Work is ongoing to compare the metric formalism adopted to the proposed 
template family and other ways of laying out the templates, for example a `stochastic' 
template bank~\cite{StochTemplBank}. Also, this explicit parametrization makes
it easier to employ additional signal-based vetoes, such as the `chi-square 
test'~\cite{Allen:2004gu}. This will also be explored in a forthcoming work.

Since this template bank is also a faithful representation of the target signals 
considered, we expect that, for a certain mass-range, a search which coherently 
includes all three stages of the binary coalescence will bring about remarkable improvement
in the estimation of parameters of the binary. This may be especially important for 
LISA data analysis in estimating the parameters of supermassive black hole binaries. 
This is also being explored in an ongoing work~\cite{ParamEstim}.

It is worth pointing out that the family of target signals (the hybrid waveforms) 
that we have considered in this paper is not unique. One can construct alternate
families of hybrid waveforms by matching PN waveforms computed using different approximations
with NR waveforms. Also, owing to the differences in initial data and accuracy 
of numerical techniques, the NR waveforms from different simulations can also 
be slightly different. Thus, the coefficients listed in Tables~\ref{tab:polCoeffsAmpParams} 
and~\ref{tab:polCoeffsPhaseParams} have a unique meaning only related to this 
particular family of target waveforms. But we expect that the general parametrization
that we propose  in this paper will hold for the whole family of non-spinning black 
hole coalescence waveforms from quasi-circular inspiral. As we have mentioned in the 
Introduction, the purpose of this paper is to explicitly prescribe a general procedure 
to construct interpolated template banks using parametrized waveforms which mimics actual 
signals from binary black hole coalescence (as predicted by numerical relativity and 
analytical methods). 

Nevertheless, it may be noted that most of the PN waveforms constructed using different 
approximations are known to be very close to each other (see, for example,~\cite{DIS01}). 
Also, we expect that NR waveforms from different simulations will converge as the 
accuracy of numerical simulations improves (see, for example,~\cite{Baker:2007fb}). 
Thus, since different families of PN and NR waveforms, which are the `ingredients' 
for constructing our target signals, are very close to each other, we expect that the 
phenomenological waveform family proposed in this paper, in its present form, will be 
sufficiently close to other families of target signals for the purpose of detecting 
these signals. As a preliminary illustration of this, we have computed the fitting 
factors of the template waveforms with a different family of hybrid waveforms (constructed 
from longer and more accurate NR waveforms), and have shown that the overlaps are indeed 
very high. This will be explored in detail in a forthcoming work.   

Also, we remind the reader that this paper consider only the leading harmonic of the 
GW signal ($\ell=2,\ m=\pm 2$). We expect that the contribution from the higher harmonics
become important for high mass ratios, which will be investigated in a forthcoming work.


\acknowledgments The authors thank Lisa Goggin and Steve Fairhurst for
help in computing the ring-down horizon distance. We also thank an
anonymous LSC reviewer for valuable comments.
The NR computations were performed with the Damiana, Belladonna and
Peyote clusters of the Albert Einstein Institute, the Doppler and
Kepler clusters at the University of Jena, as well as at LRZ Munich
and HLRS, Stuttgart. This work was supported in part by DFG grant
SFB/Transregio~7 ``Gravitational Wave Astronomy''.  The Jena group
thanks the DEISA Consortium (co-funded by the EU, FP6 project 508830),
for support within the DEISA Extreme Computing Initiative
(www.deisa.org).  AMS gratefully acknowledges the support of the
Spanish Ministerio de Educaci\'on y Ciencia research projects
FPA-2007-60220, HA2007-0042, the Conselleria D'Economia Hisenda i Innovacio of the
Government of the Balearic Islands, and the Albert Einstein Institute
and the University of Jena for hospitality.  SH thanks the University
of the Balearic Islands for hospitality, PD thanks the Albert Einstein
Institute for hospitality. YC acknowledges support from the Alexander
von Humboldt Foundation, through the Sofja Kovalevskaja Programme. The
PN waveforms were generated using the LSC Algorithms Library (LAL),
and numerical data-analysis calculations were performed with the aid
of the Merlin, Morgane and Zeus clusters of the Albert Einstein
Institute.

\appendix

\section{Calculation of the fitting factors}
\label{app:fitfactor}

In order to find the fitting factor of our phenomenological bank to a hybrid waveform, as well 
as the  best-matched parameters $(\balpha_{\rm max},\bbeta_{\rm max})$, we need to perform a 
maximization of the overlap  $\mathcal{M}(\balpha,\bbeta)$ in a 12-dimensional 
space, which seems a challenging task at first sight, especially due to the oscillatory nature 
of the dependence of $\mathcal{M}(\balpha,\bbeta)$ on the components of $\bbeta$.   However, 
due to the very high fitting factor, as well as the linear dependence of $\Psi_{\rm eff}(\bbeta; f)$ 
on $\bbeta$, we have been able to design an analytic approximation to $\mathcal{M}(\balpha,\bbeta)$ 
that is highly accurate and can be maximized over $\bbeta$ analytically.  In describing this 
approximation,  we also include $\varphi_0$ and $t_0$ in $\bbeta$, forming an 8-dimensional vector. 

For a target hybrid waveform 
\begin{equation}
\tilde h(f)  = A(f) \, e^{\rmi \Psi(f)}\,,
\end{equation}
and a phenomenological template
\begin{equation}
u(f) =  A_{\rm eff}({\balpha;f}) \, e^{\rmi \Psi_{\rm eff}(\bbeta;f)}\,,
\end{equation}
the overlap $\mathcal{M}(\balpha,\bbeta)$ can be broken into a product of two terms,
\begin{equation}
\mathcal{M}(\balpha,\bbeta) = \mathcal{M}_{\rm A}(\balpha) \,  \mathcal{M}_{\rm P}(\balpha,\bbeta)
\end{equation}
with
\begin{equation}
\mathcal{M}_{\rm A}(\balpha) \equiv
\frac{1}{a}{\int_0^{\infty} \frac{A_{\rm eff}(\balpha; f) A(f)}{S_h(f)}\df}
\end{equation} 
and
\begin{equation}
\mathcal{M}_{\rm P}(\balpha,\bbeta) \equiv 
\frac{1}{b}
\int_0^{\infty} \frac{A_{\rm eff}(\balpha; f) A(f) \cos [\Delta \Psi(f)]}{S_h(f)}\df
\end{equation} 
where
\begin{equation}
\Delta \Psi(f) \equiv \Psi(f)-\Psi_{\rm eff}(\bbeta; f)\,.
\end{equation}
In the above expressions, the normalization constants $a$ and $b$ are defined by
\begin{equation}
a^2 \equiv {\int_0^{\infty} \frac{A^2(f)}{S_h(f)}\df \int_0^{\infty} \frac{A_{\rm eff}^2(\balpha; f)}{S_h(f)}}\df,
\end{equation} 
and
\begin{equation}
b \equiv {
\int_0^{\infty} \frac{A_{\rm eff}(\balpha; f) A(f) }{S_h(f)}\df
}\,.
\end{equation} 

If the phase difference $\Delta\Psi(f)$ is small, we can approximate 
$\cos\Delta\Psi \approx  1 - \Delta\Psi^2/2$, and rewrite  $\mathcal{M}_{\rm P}$ as
\begin{equation}
\mathcal{M}_{\rm P} \approx 
\mathcal{M}_{\rm P}'
\equiv
1-\frac{1}{2\,b} \int_0^{\infty} \frac{A_{\rm eff}(\balpha; f) A(f) [\Delta\Psi(f)]^2}{S_h(f)}\df.
\end{equation}
Since $\Psi_{\rm eff}(\bbeta; f)$ is  a linear function in $\bbeta$,  minimizing $\mathcal{M}'_{\rm P}$ 
becomes a least-square fit with a weighting function 
\begin{equation}
\mu(f) \equiv \frac{A_{\rm eff}(\balpha; f) A(f)}{S_h(f)}.
\end{equation}
More specifically, writing $\Psi_{\rm eff}(\bbeta;f)$ as in Eq.(\ref{eq:phenWavePhase}), \ie,
\begin{equation}
\Psi_{\rm eff}(\bbeta;f) = \sum_j \psi_j \, f^{(5-j)/3}\,,
\end{equation}
we have
\begin{equation}
1-\mathcal{M}'_{\rm P} = \frac{1}{2} \left[
\bbeta \, \matA \, \bbeta^{\rm T} - 2 \, \vecB \, \bbeta^{\rm T} + D
\right]\,,
\end{equation}
where we have defined a matrix $\matA$, a vector $\vecB$ and a scalar constant $D$, such that
\begin{eqnarray}
A_{ij}  & \equiv & \frac {1}{b} {\int_0^{\infty}   f^{(10-i-j)/3} \, \mu(f)\, \df}\,, \nonumber \\
B_{j}   & \equiv & \frac {1}{b} {\int_0^{\infty}   f^{(5-j)/3}\Psi(f) \, \mu(f) \, \df}\,, \nonumber \\
D       & \equiv & \frac {1}{b} {\int_0^{\infty}  \Psi^2(f) \, \mu(f) \, \df}\,.
\end{eqnarray}
The maximum of $\mathcal{M}'_{\rm P}$ is then equal to 
\begin{equation}
\mathcal{M}'_{\rm P\,max} = 1-\frac{1}{2}\left[D- \vecB \, \matA^{-1}\, \vecB \right]\,,
\end{equation}
reached at
\begin{equation}
\bbeta_{\rm max} = \vecB\, \matA^{-1}.
\end{equation}
As a consequence, for each $\balpha$, we are able to maximize $\mathcal{M}_{\rm P}(\balpha,\bbeta)$, 
and hence $\mathcal{M}(\balpha,\bbeta)$,  over $\bbeta$ analytically.  The original 12-dimensional 
maximization is then converted to a 4-dimensional maximization, only over the amplitude parameters, 
on which the overlap depends in a non-oscillatory way. 


\section{Computing the horizon distance}
\label{app:horDist}

Here we describe how we compute the horizon distance of different searches
discussed in Section~\ref{sec:range}. An alternative way of computing the 
horizon distance can be found in Ref.~\cite{Ajith:2007xh}.

\subsection{Search using post-Newtonian templates}

In the restricted post-Newtonian approximation, the Fourier transform of the gravitational
signal from an optimally-oriented binary located at an effective distance $d$ can be 
written in the following way:
\be
h(f) = \frac{M^{5/6}}{d\,\pi^{2/3}}
\left(\frac{5\,\eta}{24}\right)^{1/2}\,f^{-7/6}\,e^{\rmi[2 \pi f t_0 - \varphi_0 + \psi(f) - \pi/4]}
\label{eq:hOfFPNFreqDom}
\ee
where $M$ is the total mass, $\eta$ is the symmetric mass ratio, $t_0$ is the time 
of arrival and $\varphi_0$ is the initial phase. The phase $\psi(f)$ is computed using the 
stationary phase approximation.  

The optimal SNR in detecting a known signal $h$ buried in the noise is given by 
\be
\rho_{\rm opt} = 2 \left[\int_0^{\infty} \df\, \frac{h(f)^2}{S_h(f)} \right]^{1/2},
\ee
where $S_h(f)$ is the one-sided PSD of the noise. The optimal SNR in detecting the signal
given in Eq.(\ref{eq:hOfFPNFreqDom}) can thus be computed as: 
\be
\rho_{\rm opt} = \frac{M^{5/6}}{d\,\pi^{2/3}} \left( \frac{5\,\eta}{6} \right)^{1/2}
\left[ \int_{f_{\rm low}}^{f_{\rm upp}}  \df \frac{f^{-7/3}}{S_h(f)} \right]^{1/2}\,,
\ee
where $f_{\rm low}$ is the low-frequency cutoff of the detector noise and $f_{\rm upp}$ 
is upper frequency cutoff of the template waveform. The effective distance 
to a binary which can produce an optimal SNR $\rho_{\rm opt}$ can be computed by inverting
the above equation. 

The standard post-Newtonian waveforms are truncated at $f_{\rm upp} = f_{\rm ISCO}$, 
where $f_{\rm ISCO} = (6^{3/2} \pi M)^{-1}$ is the GW frequency corresponding to the 
innermost stable circular orbit (ISCO) of the Schwarzschild geometry with mass 
equal to the total mass $M$ of the binary. The effective one body (EOB) waveforms are truncated 
at $f_{\rm upp} = f_{\rm LR}$, where $f_{\rm LR} = (3^{3/2} \pi M)^{-1}$ is the GW 
frequency corresponding to the light ring of the Schwarzschild geometry with mass $M$. 
Both of these quantities are computed assuming the test particle limit. It may be noted
that, for the EOB waveforms, an analytical Fourier domain representation is not available.
They cannot be expressed in the form given in Eq.(\ref{eq:hOfFPNFreqDom}). But for the 
purpose of the estimation of the horizon distance, these formulas give a reasonable 
approximation. 
 
\subsection{Search using ring down templates}

The ring down portion of the GW signal from a coalescing binary, considering 
only the fundamental quasi-normal mode, corresponds to a damped sinusoid. This can be written as
\cite{Echevarria:1988}
\bea
\label{gw_ringdown}
h_{\rm ring} (t) &=&  A_{\rm ring} \exp \left[-\frac{\pi f_\qnr (t-t_0)}{Q} \right]
\nonumber \\
& & \times\cos \left( -2 \pi f_\qnr(t-t_0) + \varphi_0 \right) \, ,
\eea
where $A_{\rm ring}$ is the amplitude,  $t_0$ is the start time of the 
ring down, $\varphi_0$ the initial phase, $M$ is the  mass of final black hole,
$f_\qnr$ and $Q$ are the central frequency and the quality factor of the
ringing. For the fundamental mode, a good fit to the frequency $f_\qnr$ and 
quality factor $Q$, within an accuracy of  5\%, is given by
\begin{eqnarray}
\label{f_qnr}
f_\qnr &\approx& [ 1- 0.63(1-a)^{3/10} ] \frac{1}{2 \pi M}\, ,\\ 
\label{Q}
Q &\approx& 2 (1-a)^{-9/20} \, ,
\end{eqnarray}
where $aM^2$ is the spin angular momentum, and $a$ is the Kerr parameter \cite{Echevarria:1988}.

To compute the optimal SNR in detecting this signal present in the data, 
we proceed as in \cite{Flanagan:1997sx}, assuming that for $t<t_0$,  $h_{\rm ring}(t)$ 
is identical to $t>t_0$ except for the sign in the exponential, and dividing by a 
correcting factor of $\sqrt{2}$ in amplitude to compensate for the doubling of power:
\bea
\label{gw_ringdown2}
\bar h_{\rm ring} (t) &=&  \frac{A_{\rm ring}}{\sqrt{2}} \exp \left[-\frac{\pi f_\qnr \, |t-t_0|}{Q} \right]
\nonumber \\
& & \times\cos \left( -2 \pi f_\qnr(t-t_0) + \varphi_0 \right) \, .
\eea
Its  Fourier transform then becomes
\bea
\label{gw_ringdown2f}
 {\tilde {\bar h}}_{\rm ring} (f) &=& \frac{A_{\rm ring} \, f_\qnr }{\sqrt{2}\pi \, Q} \, e^{\rmi 2\pi f t_0} \,
\left(\frac{e^{\rmi \varphi_0}}{ {g}^2 +4\,(f-f_\qnr)^2 } \right. \nonumber \\
& + & \left. \frac{e^{-\rmi \varphi_0}}{ {g}^2 +4\,(f+f_\qnr)^2 }\right)\,,
\eea
where $g=f_\qnr/Q$. 

In general, it is not easy to estimate $A_{\rm ring}$, or the two polarization amplitudes; 
they depend upon the detailed evolution of the merger epoch, as well as variables such as 
the orientation of the final  merged remnant. A reasonable hypothesis \cite{Goggin:2005,Hughes:2001ya,Fryer:2001zw}
is that their ratio follows the ratio of the inspiral polarization amplitudes. With this 
hypothesis, the overall amplitude of the signal from an optimally located and  oriented binary,
requiring that the ring down radiate some fraction $\epsilon$ of the system's total mass, 
becomes
\be
A_{\rm ring}^{\rm opt}= \sqrt{\frac{5\epsilon }{4\pi}}\frac{M}{d}
\frac{2}{\sqrt{M f_\qnr Q F(Q)  }}
\ee
where $F(Q)=1+\frac{7}{24 Q^2}$ and $d$ is the distance to the source. The optimal 
SNR $\rho$ can now be computed as
\be
\rho_{\rm opt} = 2\, \left[\int_{f_{\rm low}}^{f_{\rm upp}}  
\df \frac{\vert {\tilde {\bar h}}_{\rm ring}\vert^2}{S_h(f)} \right]^{1/2}\,,
\ee
where $f_{\rm low}$ and $f_{\rm upp}$ are the lower and upper cutoff frequencies of 
the detector noise. As in the previous case, the horizon distance can be computed by inverting 
this equation.

\subsection{Search using the template family proposed in this paper}

The phenomenological waveforms in the frequency domain are given in Eqs.(\ref{eq:phenWave}--
\ref{eq:phenWaveAmpConst}). The optimal SNR in detecting this signal can be computed as: 
\ber
\rho_{\rm opt} & = & \frac{M^{5/6}\,\fmerg^{-7/6}}{d\,\pi^{2/3}} \left( \frac{5\,\eta}{6} \right)^{1/2}
\left[ \int_{f_{\rm low}}^{\fmerg}  \df \frac{(f/\fmerg)^{-7/3}}{S_h(f)} \right. \nonumber \\
& + & \int_{\fmerg}^{\fring} \df \frac{(f/\fmerg)^{-4/3}}{S_h(f)} \nonumber \\
& + & w^2 \left. \int_{\fring}^{\fcut} \df \frac{{\cal L}^2(f,\fring,\sigma)}{S_h(f)} \right]^{1/2}, 
\eer
where ${\cal L}(f,\fring,\sigma)$ is defined in Eq.(\ref{eq:Lorenzian}), and $\fmerg, \fring, \fcut$ 
and $\sigma$ are given by Eq.(\ref{eq:reparametr}).

This equation can be inverted to calculate the effective distance to the optimally-oriented 
binary which can produce an optimal SNR $\rho_{\rm opt}$.

\bibliography{template}

\end{document}